\documentclass[11pt]{article}
\usepackage{latexsym}
\usepackage[mathscr]{eucal}
\usepackage{amscd}
\usepackage{amssymb}
\usepackage{verbatim}

\newcommand\beq{\begin{eqnarray}}
\newcommand\eeq{\end{eqnarray}}
\newcommand\ben{\begin{enumerate}}
\newcommand\een{\end{enumerate}}
\renewcommand{\a}{\alpha}
\renewcommand{\b}{\beta}
\newcommand{\der}{\partial_x}

\newcommand{\tzeta}{\widetilde \zeta}
\newcommand{\tR}{\widetilde R}
\newcommand{\tSi}{\widetilde \Sigma}
\newcommand{\tSigma}{\widetilde \Sigma}

\newcommand{\tN}{\widetilde N}
\newcommand{\tL}{\widetilde L}
\newcommand{\ttime}{\widetilde t}
\newcommand{\tn}{\widetilde n}
\newcommand{\wh}{\widetilde h}
\newcommand{\tna}{\widetilde \nabla}
\newcommand{\tg}{\widetilde g}
\newcommand{\tD}{\widetilde \Delta}
\newcommand{\tW}{\widetilde W}

\newcommand{\tS}{\widetilde S}
\newcommand{\tP}{\widetilde P}
\newcommand{\tWeyl}{\widetilde C}
\newcommand{\ts}{\tilde \sigma}% This isn't tilde s
\newcommand{\tsigma}{\tilde \sigma}

\newcommand{\tb}{\widetilde b}

\newcommand{\tnabla}{\widetilde \nabla}
\newcommand{\tH}{\widetilde H}

\newcommand{\qed}{\hfill $\Box$ \medskip}
\newcommand{\re}{{\mathbb R}}

\newcommand{\zz}{{\mathbb Z}}

\newcommand{\pd}{\partial}
\newcommand{\cM}{{\cal M}}

\newcommand{\cH}{{\cal H}}
\newcommand{\cW}{{\cal W}}
\newcommand{\cU}{{\cal U}}
\newcommand{\bK}{K}
\newcommand{\bD}{{\mathbb D}}
\newcommand{\fG}{G}

\newcommand{\fg}{{\mathfrak g}}

\makeatletter
\renewcommand{\section}{\@startsection{section}{1}{0mm}%
{-0.5\baselineskip}{1 pt}{\normalfont\large\bfseries}}
\renewcommand{\subsection}{\@startsection{subsection}{2}{0mm}%
{-0.1\baselineskip}{1 pt}{\normalfont\bfseries}}
\renewcommand{\subsubsection}{\@startsection{subsubsection}{3}{0mm}%
{-0.1\baselineskip}{1 pt}{\normalfont\bfseries}}
\renewcommand{\paragraph}{\@startsection{paragraph}{4}{0mm}%
{-0.1\baselineskip}{1 pt}{\normalfont\itshape}}
\renewcommand{\subparagraph}{\@startsection{subparagraph}{5}{0mm}%
{-0.1\baselineskip}{1 pt}{\normalfont}}
\makeatother

\renewcommand{\thesection}{\Roman{section}}
\renewcommand{\thesubsection}{\thesection.\arabic{subsection}}

\title{On the Geometry and Mass of Static, Asymptotically AdS Spacetimes,
and the Uniqueness of the AdS Soliton}

\author{G.J.\ Galloway${}^{a}$\footnote{galloway@math.miami.edu},
S.\ Surya${}^{b,c}$\footnote{ssurya@phys.ualberta.ca},
E.\ Woolgar${}^{c,b}$\footnote{ewoolgar@math.ualberta.ca}
\\
\\
${}^a$Dept.\ of Mathematics, University of Miami\\
Coral Gables, FL 33124, USA
\\
\\
${}^b$Theoretical Physics Institute, Dept.\ of Physics\\
University of Alberta, Edmonton, AB, Canada T6G 2J1
\\
\\
${}^c$Dept.\ of Mathematical and Statistical Sciences\\
University of Alberta, Edmonton, AB, Canada T6G 2G1}

\date{}

\begin{document}

\maketitle
\begin{abstract}
We prove two theorems, announced in hep-th/0108170, for static
spacetimes that solve Einstein's equation with negative
cosmological constant. The first is a general structure theorem
for spacetimes obeying a certain convexity condition near
infinity, analogous to the structure theorems of Cheeger and
Gromoll for manifolds of non-negative Ricci curvature. For
spacetimes with Ricci-flat conformal boundary, the convexity
condition is associated with negative mass. The second theorem is
a uniqueness theorem for the negative mass AdS soliton spacetime.
This result lends support to the new positive mass conjecture due
to Horowitz and Myers which states that the unique lowest mass
solution which asymptotes to the AdS soliton is the soliton
itself. This conjecture was motivated by a nonsupersymmetric
version of the AdS/CFT correspondence. Our results add to the
growing body of rigorous mathematical results inspired by the
AdS/CFT correspondence conjecture. Our techniques exploit a
special geometric feature which the universal cover of the soliton
spacetime shares with familiar ``ground state'' spacetimes such as
Minkowski spacetime, namely, the presence of a null line, or
complete achronal null geodesic, and the totally geodesic null
hypersurface that it determines. En route, we provide an analysis
of the boundary data at conformal infinity for the Lorentzian
signature static Einstein equations, in the spirit of the
Fefferman-Graham analysis for the Riemannian signature case. This
leads us to generalize to arbitrary dimension a mass definition
for static asymptotically AdS spacetimes given by Chru\'sciel and
Simon. We prove equivalence of this mass definition with those of
Ashtekar-Magnon and Hawking-Horowitz.
\end{abstract}
\vspace{0.1 cm}

\section{Introduction}
\renewcommand{\theequation}{\thesection.\arabic{equation}}
\setcounter{equation}{0}

\noindent There has been considerable interest in the last few
years in the AdS/CFT correspondence first proposed in
\cite{adscft,witten98,gkp}. This correspondence equates the string
partition function on an anti-de Sitter (AdS) background to that
of a conformal field theory on the AdS boundary-at-infinity (the
Penrose conformal boundary, scri), and is an explicit realisation
of the holographic principle suggested in \cite{'tHooft,susskind}.
In particular the large $N$ limit of the conformal field theory
corresponds to the low energy limit of string theory, i.e.,
classical supergravity. In this limit, classical properties of a
spacetime have definite interpretations in the gauge theory (see
\cite{review} for example).

Of particular interest is the question of what the positivity of
gravitational energy means in the conformal field theory. Related
to this is the role of ``ground'' states, namely, the lowest
energy configurations classically allowed, satisfying certain
physically reasonable conditions. In asymptotically flat space,
the celebrated positive energy theorem \cite{schyau,witten81}
tells us that Minkowski space is the unique lowest energy solution
provided that the local energy density is non-negative and that
there are no naked singularities. In the presence of a negative
cosmological constant, the analogous ground state is AdS
spacetime. Both Minkowski spacetime and AdS are regular, static,
supersymmetric and of constant curvature. For each of these
familiar ground states there is an associated uniqueness theorem
\cite{lich,bgh,CS,CH} showing that the spacetime is unique in the
class of regular, stationary, vacuum solutions when the
appropriate fall-off conditions are satisfied.

The standard conditions for an $(n+1)$-dimensional spacetime to be
asymptotically AdS include a specification of the topology of the
$n$-dimensional conformal boundary, i.e., that it be
$S^{n-1}\times \re$ \cite{AM}. When this topological restriction
is relaxed, black hole spacetimes which are asymptotically locally
AdS and which have nonspherical horizons are known to exist\footnote
{In what follows, the term {\it locally Anti-de Sitter spacetime}
means ``constant negative curvature spacetime'', while {\it
Anti-de Sitter (AdS) spacetime} refers to the unique, geodesically
complete, simply connected, constant negative curvature spacetime.
An asymptotically AdS spacetime has spherical scri, but one that
is merely asymptotically locally AdS need not have spherical
scri.}
(for a review, see \cite{mann}). The timelike conformal infinities
of these black holes have nonspherical cross-sections \cite{tc}.
However, a conformal boundary with cross-sections of nonspherical
topology cannot also serve as the conformal boundary of a
nonsingular locally AdS spacetime. Thus, AdS or its topological
modifications cannot be appropriate ground states for such black
hole spacetimes.

Thus we are led to entertain the somewhat radical proposition that
ground states for spacetimes with nonspherical scri may not be
conformally flat, and thus not massless according to familiar mass
formulae such as Ashtekar-Magnon \cite{AM}. For the case of
toroidal scri, a candidate ground state has been proposed by
Horowitz and Myers \cite{hm}, which they dubbed the ``AdS
soliton''. In $n+1 \geq 4$ spacetime dimensions,\footnote
{In $2+1$ spacetime dimensions, this soliton is identical to
$3$-dimensional AdS.}
it is a globally static Einstein spacetime with cosmological
constant $\Lambda<0$ and has the form
\begin{equation}
ds^2 = -r^2 dt^2 +  \frac{dr^2}{V(r)} + V(r) d\phi^2 + r^2
\sum\limits_{i=1}^{n-2}(d\theta^i)^2\quad . \label{Intro1}
\end{equation}
Here, $V(r)=\frac{r^2}{\ell^2}(1- \frac{r_0^n}{r^n})$, with
$\ell^2=\frac{n(n-1)}{-2\Lambda}$ and $r_0$ is a constant. The
solution is nonsingular provided $\phi$ is periodic with period
$\beta_0=\frac{4\pi\ell^2}{nr_0}$. The periods of the
$\theta^i$ are arbitrary. The soliton is asymptotically locally
Anti-de Sitter with boundary-at-infinity (scri) foliated by
spacelike $(n-1)$-tori. Spacetime itself, when conformally
completed, has constant time slices which are topologically the
product of an $(n-2)$-torus and a disk (in $3+1$ dimensions, it is
therefore a solid torus). Moreover, the soliton is neither of
constant curvature nor supersymmetric, but it is known to have
minimal energy under small metric perturbations \cite{hm,cm}.

Curiously, the AdS soliton has negative mass relative to the
natural choice that associates the zero of energy with conformal
flatness. The mass $E_0$ depends on the period $\beta_0$ of the
$\phi$ coordinate according to
\begin{equation}
E_0=\frac{-V_{n-2}\ell^{n-1}}{16\pi G_{n+1}\beta_0^{n-1}}
\left ( \frac{4\pi}{n}\right )^n \quad ,\label{Intro2}
\end{equation}
where $V_{n-2}$ is the product of the periods $\beta_i$ of the $\theta^i$,
these being arbitrary, and $G_{n+1}$ is Newton's constant in the
$(n+1)$-dimensional spacetime.
A simple argument may suggest that by rescaling the
parameters of a negative mass spacetime, one could get a spacetime
of even lower mass. However, for the AdS soliton, the proposed
rescaling is an isometry, and does not change the mass
\cite{GSW1}.
\footnote{The black hole solutions of Lemos \cite{Lemos} are also
mutually isometric under this scaling, and a similar scaling works
for the $5$-dimensional nilgeometry and solvegeometry black holes
of \cite{CW}.}
Thus, a negative mass ground state, in this case, need not be a
contradiction after all.
Page \cite{Page} has introduced a scale-invariant mass
\begin{equation}
\epsilon := E_0V_{n-1}:=E_0V_{n-2}\beta_0=-C\left ( \frac{<\beta>}
{\beta_0}\right )^n\quad ,\label{Intro3}
\end{equation}
where $C$ is a constant that depends on $n$ but is invariant under
rescalings of the periods, and
$<\beta>:=(\beta_0\beta_1\dots\beta_{n-2})^{1/(n-1)}$
is the geometric mean of these periods, including $\beta_0$.

Horowitz and Myers found that the negative mass of the AdS
soliton has a natural interpretation as the Casimir energy of a
nonsupersymmetric gauge theory on the conformal boundary. If a
nonsupersymmetric version of the AdS/CFT conjecture is to hold,
as is generally hoped, then this would indicate that the soliton
is the lowest energy solution with these boundary conditions.
This led them to postulate a new positive energy conjecture, that the
soliton is the unique lowest mass solution for all spacetimes in
its asymptotic class. The validity of this conjecture is thus an
important test of the nonsupersymmetric version of the AdS/CFT
correspondence.\footnote
{Page (\cite{Page} points out that, given $n-1$ distinct positive
numbers $b_{\mu}$, there are $n-1$ distinct solitons for which
these numbers serve as boundary data specifying the periods of
the coordinates, depending on which one of them is chosen to equal
the period $\beta_0$ of $\phi$, and that the invariant energy can
be minimized only if $\beta_0 ={\rm Min}_{\mu} \{b_{\mu}\}$.}

A preliminary indication that the soliton is the appropriate
ground state comes from a semi-classical thermodynamic analysis of
Ricci flat black holes in the background of the AdS soliton
\cite{ssw}. An examination of the thermodynamics of the spherical
AdS black hole showed that there is a phase transition that takes
place between the black hole and the appropriate ground state,
namely AdS spacetime \cite{hawpage}. This phase transition was
interpreted as a confinement/deconfinement transition in the
associated large $N$ gauge theory on the boundary \cite{witten98}.
Taking a cue from this, it was shown that a phase transition also
occurs between the toroidal black hole and the AdS soliton, and
can also be interpreted as a confinement/deconfinement transition
in the boundary field theory, in much the same way \cite{ssw}.
Indeed the choice of ground state is crucial to see such a phase
transition. Earlier analysis which identified the {\it locally}
AdS spacetime with toroidal boundary (which is singular) as the
ground state did not give rise to such a phase transition.

In support of the Horowitz-Myers new positive energy conjecture,
we prove a uniqueness theorem for the AdS soliton, singling it out
as the only suitable ground state in a large class of negative
mass spacetimes obeying certain boundary conditions. Our results
are similar in spirit to those of \cite{lich,bgh}, but relate to
asymptotically locally AdS spacetime with Ricci flat scri. The key
elements of the proof are, briefly, (a) the use of negative mass
and certain asymptotic conditions (related to the convexity of
constant lapse hypersurfaces near scri) to establish the existence
of null lines in the universal covering spacetime and (b) the
construction, using the null splitting theorem due to Galloway
(\cite{gjg}, quoted below as Theorem I.1), of a foliation of
spacetime based on totally geodesic null hypersurfaces.

A \emph{null line} in spacetime is an inextendible null geodesic which is
globally achronal, i.e., for which no two points can be joined by a timelike
curve. (Hence, each segment of a null line is maximal with respect to the
Lorentzian distance function.) Arguments involving null lines have arisen
in numerous situations, such as the Hawking-Penrose singularity theorems
\cite{HE}, results on topological censorship (\cite{tc}, and references cited
therein), and the Penrose-Sorkin-Woolgar approach to the positive mass theorem
\cite{PSW,Wool} and related results on gravitational time delay \cite{GW}.
It will be convenient for the purposes of the present paper to require null
lines to be not only inextendible, but geodesically complete.

For the reader's convenience, we quote here the null splitting theorem:

\noindent {\bf Theorem I.1 (Galloway \cite{gjg}).} {\sl If a null
geodesically complete spacetime obeys $R_{ab}X^aX^b$ $\ge 0$ for all
null vectors $X^a$ and also contains a null line $\eta$, then $\eta$
lies in a smooth, achronal, edgeless, \emph{totally geodesic} null
hypersurface ${\cal H}$.}

\noindent {\bf Remark I.2.} (a) The spacetimes considered herein
will be vacuum (with negative cosmological constant), so they obey
$R_{ab}X^aX^b=0$ for all null vectors $X^a$. (b) Because ${\cal
H}$ is totally geodesic, the tangent vector field $n^a$ to the null
geodesic generators of ${\cal H}$, when suitably scaled, is a parallel
vector field along $\cal H$, $X^a\nabla_a n^b=0$ for all tangent vectors
$X$ to $\cal H$. We will make use of this fact in Section III.

In Section \ref{prelim.sec}, we consider a boundary value problem
for asymptotically locally AdS spacetimes with Ricci flat
conformal boundary. We draw on a formalism of Chru\'sciel and
Simon \cite{CS} for static, asymptotically AdS solutions, but
while they restrict to $4$ spacetime dimensions, we work in $n+1$
spacetime dimensions. To discuss boundary conditions, and in
particular to relate the sign of the mass to data on the Penrose
conformal boundary, we express the extrinsic geometry of
hypersurfaces approaching scri as an expansion in a certain
coordinate distance from scri, and use this to expand the scalar
curvature of the conformal metric on constant time slices. Similar
analyses appear in \cite{HT, FG, HS, CS, Anderson}. We find that
in the static spacetime setting with Ricci flat boundary a power
series expansion suffices for both even and odd dimension $n$,
i.e., no log terms arise in the expansion. The only free data is
the induced boundary conformal metric and its $n^{\rm th}$
``radial'' derivative, which we relate to the mass. We find it
convenient to use the Chru\'sciel-Simon mass definition,
generalized to arbitrary dimension. In Section II.3, we prove
equivalence of this mass and the Ashtekar-Magnon mass. A proof of
equivalence to the Hawking-Horowitz mass is consigned to Appendix
A. Given these equivalences, then equivalence to other common mass
definitions (esp.\ that of Abbott and Deser \cite{AD}) follows
from existing results \cite{HH}.

In Section \ref{structure.sec}, we present the first of our main
theorems, a structure theorem. Here we show, roughly speaking,
that given a certain convexity condition near infinity, then in
the universal covering space of a constant time slice in the
conformal spacetime, the noncompact directions split off from the
compact directions, and are flat. This result is analogous to a
structure theorem of Cheeger and Gromoll \cite{CG}. The proof
works by using our asymptotic conditions to establish the presence
of a line in the universal cover. This line can be lifted to a
null line in the universal cover of the physical spacetime; this
is proved in Appendix B. By the null splitting theorem I.1,
spacetimes with null lines have a special geometry, and this
yields our structure result. By further imposing certain
topological restrictions, we are then led to a uniqueness theorem
for the soliton which we prove in Section \ref{uniqueness.sec}.
The assumption of negative mass in the uniqueness theorem is used,
via the results of Section II, to show that the aforementioned
convexity condition holds \emph{in the mean}, thereby permitting a
weakening of the convexity condition (cf.\ the discussion of
Conditions (C) and (S) in Section III.1). We very briefly mention
some extensions of our results in Section \ref{concluding.sec}.

Of related interest, Anderson \cite{Anderson} has proved uniqueness
for $4$-dimensional hyperbolic (thus, Riemannian) metrics, provided
certain coefficients in the expansion of the conformal metric are fixed
on the conformal boundary. Also, Kiem and Park \cite{KP} have shown
uniqueness of the soliton but only under very strong assumptions,
among which, for example, is the structure theorem that we will prove
in Section III.

The main theorems proved in Sections III and IV were announced in
\cite{GSW1}. Herein, we provide explicit, detailed proofs and
associated analyses and lay a basis for the further future work
\cite{GSW2} briefly touched upon in Section \ref{concluding.sec}.
Throughout, the spacetime dimension is $n+1$.

\section{The Boundary Value Problem} \label{prelim.sec}
\renewcommand{\theequation}{\thesubsection.\arabic{equation}}
\setcounter{equation}{0}

\noindent
The static Einstein equations in the asymptotically locally de Sitter
setting form a highly nonlinear elliptic, asymptotically degenerate,
system of equations, and
it is not clear {\it a priori} that the AdS soliton would be a
unique solution even if all the necessary boundary data were
specified. We write out the field equations with respect to the
physical metric and also with respect to a relevant conformally
related metric in Section II.1. In Section II.2, we find that the
free data on the (Ricci flat) conformal boundary are
the induced metric and its normal derivatives of order $n$; the
latter also determine the mass of the spacetime. It is convenient
to use a mass definition based on that of Chru\'sciel and Simon \cite{CS},
so in Section II.3 we prove equivalence of this mass to the
familiar Ashtekar-Magnon mass. A feature of our uniqueness theorem
is that we will not need to specify all the free data on the
conformal boundary to obtain uniqueness. Apart from the induced
boundary metric, we will specify only that
the sign of the mass is negative, though we must pay a price by
requiring further topological assumptions and an assumption on the
extrinsic geometry of constant lapse surfaces near infinity
(cf.\ Section III.1).

\subsection{The Field Equations} \label{FE}
\setcounter{equation}{0}

\noindent
We consider $(n+1)$-dimensional, $n\ge 2$,
static spacetimes $(\cM, g)$,
\begin{eqnarray}
\cM &=& \mathbb R\times \Sigma\quad , \nonumber \\
g &=& -N^2dt^2 \oplus h \quad ,
\label{FE1}
\end{eqnarray}
where $h$ is the induced metric on $\Sigma$ and $N$ is the lapse,
such that the triple $(\Sigma, h, N)$ is $C^k$ ($k\ge 2$)
conformally compactifiable. Thus, $\Sigma$ is the interior of a
smooth compact manifold with boundary
$\tSigma=\Sigma\cup\partial\tSigma$ such that
\begin{enumerate}
\item[(a)] $N^{-1}$ extends to a $C^k$ function $\tN$ on $\tSigma$,
with $\tN|_{\partial \tSigma}=0$ and $d\tN|_{\partial \tSigma} \neq 0$
pointwise, and
\item[(b)] $N^{-2}h$ extends to a $C^k$ Riemannian metric $\wh$
known as the {\it Fermat} (or \emph{optical}) metric
on $\tSigma$.
\end{enumerate}
The definition of conformally compactifiable given here precludes
the existence of internal boundaries, such black hole boundaries.
While this preclusion is not needed for the asymptotic analysis
presented in this section, it is used in Sections III and IV,
which are concerned with properties of \emph{globally} static
spacetimes. The case of black holes will be dealt with in a
forthcoming paper \cite{GSW2}.

The triplet $(\Sigma, h, N) $ obeys the static vacuum field equations
\begin{eqnarray}
R_{ab}& = & \frac{1}{N}\nabla_a\nabla_b N +\frac{2\Lambda}{n-1} h_{ab}
\label{FE2}, \\
\Delta N & = & -\frac{2\Lambda}{n-1} N,
\label{FE3}
\end{eqnarray}
where $\nabla_a$ and $R_{ab}$ are respectively the covariant
derivative and Ricci tensor on $(\Sigma, h)$, and $\Lambda <0$ is the
cosmological constant. These spacetimes are asymptotically
constant negative curvature.

These equations can be rewritten in terms of the Fermat metric
$\wh$ and associated $\tnabla_a$ and $\tR_{ab}$ as
\begin{eqnarray}
\tR_{ab}&=&\frac{-(n-1)}{\tN}\tnabla_a\tnabla_b \tN\quad ,\label{FE4}\\
\tN \tD\tN&=&\left (\frac{2\Lambda}{n-1}+n\tW\right ) \quad ,\label{FE5}
\end{eqnarray}
where
\begin{equation}
\tW:={\wh^{ab}}\tnabla_a\tN\tnabla_b\tN =\frac{1}{N^2}h^{ab}\nabla_a N\nabla_b
N\label{FE6}.
\end{equation}
A useful identity is obtained by taking the trace of (\ref{FE4}) and
combining this with (\ref{FE5}):
\begin{equation}
\tN^2 \tR +2\Lambda +n(n-1)\tW = 0\quad .\label{FE7}
\end{equation}
Solving for $\Lambda$ and reinserting into (\ref{FE5}), we obtain
\begin{equation}
\biggl(\tD + \frac{\tR}{n-1}\biggr)\tN=0\quad .\label{FE8}
\end{equation}

We will sometimes state results in terms of the triple $(\Sigma,h,N)$, but
because we assume conformal compactifiability, we will often work
with $(\tSigma,\wh,\tN)$, and therefore we often work with the system
(\ref{FE4}--\ref{FE6}) rather than (\ref{FE2}--\ref{FE3}).

%%EW: Following material useful to restore?
%%GG: But then it seems to me this should be used to prove Corollary
%II.2.3, which
%would require further changes.  Do you want to do that?  If so, I think it
%would be better to introduce this equation in Corollary II.2.3.  Also, I think
%old conventions  are being used in the second displayed equation below (with
%regard to dimension, value of \Lambda).
%%Begin Comment
\begin{comment}
It will be useful in what follows to express the Laplacian of
$\tN$ in terms of the mean curvature of constant $\tN$
hypersurfaces. The (Fermat) unit normal to such a hypersurface is
\begin{equation}
{\tn}^a=\frac{-1}{\sqrt{\tW}}\wh^{ab}{\tnabla}_b \tN\quad ,
\label{FE9}
\end{equation}
This normal is ``outward directed''; at the $\tN=0$ hypersurface,
which is a cut of scri, it points away from spacetime. From this
definition, it is easy to see that the mean curvature $\tH:=
\tnabla_a \tn^a$ of the constant $\tN$ hypersurfaces can be
expressed as
\begin{eqnarray}
\sqrt{\tW}\tH
&=&\frac{-1}{2\sqrt{\tW}}\tn^a\tnabla_a\tW-\tD\tN\nonumber\\
&=&\frac{x\tR}{n}-\frac{x^2}{2n(n-1)}\frac{\partial \tR}{\partial
x}\quad ,
\end{eqnarray}
where in the last step we have applied (\ref{FE7}) and (\ref{FE8})
and defined $x:=\tN$.
\end{comment}
%%EW: End Comment

We end the section with an identity that will be of use in our
discussion of mass and boundary conditions. We define $x:=\tN=1/N$
and let $\partial/\partial x$ be the vector dual to $dx$ under the
isomorphism defined by the Fermat metric $\wh_{ab}$. If we then
Fermat normalize this dual vector, we obtain the vector
\begin{equation}
{\tn}^a:=\frac{-1}{\sqrt{\tW}}\left ( \frac{\partial}{\partial x}
\right )^a =\frac{-1}{\sqrt{\tW}}\wh^{ab}{\tnabla}_b \tN\quad .
\label{FE9}
\end{equation}
Note that we have also reversed the sense of the vector by introducing
a minus sign. This is for convenience in the next section, where we
will use this formula in the case where $\tn^a$ will be normal to scri,
and we will want it to be the outward directed normal, pointing in
the direction of decreasing $x$. Now from the definition (\ref{FE6})
we have
\begin{eqnarray}
\frac{\partial \tW}{\partial x}=\frac{-1}{\sqrt{\tW}}\tn^a\tnabla_a\tW
&=& \frac{-2}{\sqrt{\tW}}\left ( \wh^{ab}\tnabla_a x \right )
\left ( \tn^c \tnabla_c\tnabla_b x\right )\nonumber\\
&=&\frac{-2x}{(n-1)}\tR_{ab}\tn^a\tn^b\quad , \label{FE10}
\end{eqnarray}
where in the last step we used (\ref{FE4}). If we now differentiate
(\ref{FE7}) with respect to $x=\tN$, use the results in the
left-hand side of (\ref{FE10}), and rearrange terms, we obtain
\begin{equation}
\tR_{ab}\tn^a\tn^b-\frac{1}{n}\tR
=\frac{x}{2n}\frac{\partial \tR}{\partial x}
\quad .\label{FE11}
\end{equation}

\subsection{The Boundary Conditions}
\setcounter{equation}{0}

\noindent The solution of the field equations on $\Sigma$ of course will
not be unique unless we specify some boundary data on $\partial \tSigma$,
which here is the hypersurface $x:=\tN =0$. The data we wish to
specify are the induced metric on $\partial \tSigma$ and the sign of the
mass of spacetime. The latter is related to the $n^{\rm th}$ order
$x$-derivatives of the Fermat metric coefficients at $x=0$.

Many similar analyses have appeared in the literature, among them
\cite{HT, FG, HS, CS, Anderson}. These analyses usually focus on
the issue of whether the vacuum Einstein equations admits a formal
power series solution centred at conformal infinity or whether the
power series must be supplemented by log terms. Typically, these
analyses deal with the full vacuum Einstein equations, either with
zero \cite{FG} or negative \cite{HS} cosmological constant. We
will deal with the static (thus, Lorentzian signature) Einstein
equations in dimension $n+1$, and will focus on the case of
Ricci-flat conformal boundary since this case includes the AdS
soliton. The assumption of a timelike Killing field yields a more
restrictive system of equations than the general system with no
symmetries in $n+1$ dimensions (that it is not equivalent to the
$n$-dimensional system is evident, cf.\ (\ref{FE2}) below). We
find formal power series solutions for all $n$, in agreement with
the results of \cite{HS} in dimensions $2$, $4$, and $6$.

We show below that the first $n-1$ $x$-derivatives of the Fermat
metric components at $x=0$ vanish if this metric is assumed to be
of class $C^n$ there. From what follows, it can be seen that if
the order $n$ $x$-derivatives of the Fermat metric components are
supplied at $x=0$, then all higher $x$-derivatives are determined
there, up to one order below that at which differentiability
fails. This behaviour does not depend on whether $n$ is even or
odd. When the Fermat metric is of class $C^{\infty}$ at $x=0$,
this technique yields a formal power series solution of the field
equations at $x=0$. Here we do not assume $C^{\infty}$, nor do we
concern ourselves with convergence of the power series, preferring
instead to obtain our uniqueness theorem by geometric techniques.
An important ingredient of these techniques will be Corollary
II.2.3, which relates the sign of the mass aspect (Definition
II.3.2) to the mean curvature of constant lapse surfaces near
infinity.

Near the boundary $x=0$, we may introduce coordinates $x^1 = x,
x^2,\dots,x^n$ so that the metric $\wh$ takes the form,
\begin{equation}
\label{BC2} \wh = \frac{dx^2}{\tW} + \tb_{\a\b} \,dx^{\a}dx^{\b}
\quad,
\end{equation}
where $\tb_{\a\b} =\tb_{\a\b}(x,x^{\gamma})$ is the induced metric
on the constant $x$ slice $V_x\approx
\partial\tSi$ (here and throughout this subsection, Greek indices
run from $2$ to $n$). The second fundamental form $\tH_{\a\b} =
\tH_{\a\b}(x)$ of $V_x$ is defined as $\tH_{\a\b} =
\tna_{\a}\tn_{\b}$, from which it follows that
\begin{equation}
\tH_{\alpha\beta} =\frac{-1}{2\psi}\partial_x\tb_{\alpha\beta} \quad ,
\label{BC3}
\end{equation}
where $\psi:=\tW^{-1/2}$.

By taking the projections of the field equation (\ref{FE4}) tangent
and normal to each $V_x$, we obtain
\begin{eqnarray}
\tR_{\alpha\beta}&=&\frac{(n-1)}{x\psi}\tH_{\alpha\beta}\quad ,\label{BC4}\\
\tR_{xx}&=&\frac{(n-1)}{x\psi}\partial_x\psi\quad .\label{BC5}
\end{eqnarray}

We may use the standard expression for the Ricci curvature in terms of
Christoffel symbols to expand the left-hand sides of equations
(\ref{BC4}, \ref{BC5}). Doing so, we obtain
\begin{eqnarray}
\tH_{\alpha\beta}&=&\frac{x}{n-1}\left ( {\der {\tH}}_{\alpha\beta}
+\left [ 2\tH_\alpha {}^{\gamma} \tH_{\beta\gamma}-\tH\tH_{\alpha\beta}
+{\cal R}_{\alpha\beta} -D_\alpha D_\beta \right ] \psi \right ) \quad ,
\label{BC6}\\
{\der \psi}(x) &=&\frac{x}{n-1}\psi^2\left ( {\der {\tH}} -\left [ D^2
+\tH_{\alpha\beta}\tH^{\alpha\beta}\right ]\psi\right ) \quad ,\label{BC7}
\end{eqnarray}
where $D_\alpha $ is the Levi-Cevita connection of the induced metric on
$V_x$ and ${\cal R}_{\alpha\beta}$ is its Ricci tensor.
We see from (\ref{BC6}, \ref{BC7}) (or directly from (\ref{BC4},
\ref{BC5})) that $C^2$ regularity of the Fermat metric at $x=0$
requires that
\begin{equation}
\tH_{\alpha\beta}(0)={\der \psi}(0)=0\quad .\label{BC8}
\end{equation}

If the terms in (\ref{BC6}, \ref{BC7}) are $k-1$ times
differentiable then, taking $k\ge 2$, we may apply
$\partial^{k-1}_x := \partial^{k-1}/\partial x^{k-1}$ to these equations
to obtain, for $n\neq k$, the following expressions:
\begin{eqnarray}
\partial_x^{k-1}\tH_{\alpha\beta}&=&
\frac{x}{(n-k)}\partial_x^{k-1}\left ( {\der {\tH}}_{\alpha\beta}
+\left [ 2\tH_\alpha {}^{\gamma}\tH_{\beta\gamma}-\tH\tH_{\alpha\beta}
+{\cal R}_{\alpha\beta}-D_\alpha D_\beta\right ]\psi \right )\nonumber\\
&&+\frac{(k-1)}{(n-k)} \partial_x^{k-2}\left (\left [ 2\tH_\alpha {}^{\gamma}
\tH_{\beta\gamma}-\tH\tH_{\alpha\beta}+{\cal R}_{\alpha\beta}-D_\alpha
D_\beta\right]\psi \right ) \quad , \label{BC9}\\
\partial_x^{k}\psi &=&
\frac{x}{(n-1)}\partial_x^{k-1} \left ( \psi^2{\der {\tH}} -\psi^2
\left [ D^2+\tH_{\alpha\beta}\tH^{\alpha\beta}\right ]\psi\right )
\nonumber\\
&&+\frac{(k-1)}{(n-1)}\partial_x^{k-2}\left ( \psi^2{\der {\tH}}
-\psi^2 \left [ D^2 + \tH_{\alpha\beta}\tH^{\alpha\beta}\right ]
\psi \right ) \quad .\label{BC10}
\end{eqnarray}

\noindent {\bf Proposition II.2.1.} {\sl Let $(\Sigma,h,N)$
be $C^{m+1}$ conformally compactifiable. We assume the conformal
boundary hypersurface $x=0$ to be Ricci flat. ({\it i})
If $m<n:={\rm dim}\Sigma$, the first $m$ $x$-derivatives of $\psi$
and of $\tb_{\alpha\beta}$ (equivalently, from (\ref{BC3}),
$\tH_{\alpha\beta}(x)$ and its first $m-1$ $x$-derivatives)
vanish at $x=0$. ({\it ii}) If $m\ge n$, the first $n-1$ $x$-derivatives
of $\psi$ and $\tb_{\alpha\beta}(x)$ vanish at $x=0$ and the remaining
$x$-derivatives up to order $m-1$ inclusive are completely determined
by $\tb_{\alpha\beta}(0)$ and $\tb_{\alpha\beta}^{(n)}(0)$
(equivalently, $\tH_{\alpha\beta}^{(n-1)}(0)$).}

\noindent{\bf Proof.} By assumption, the Fermat metric is
$C^{m+1}$ differentiable, which is equivalent to $C^{m+1}$
differentiability of $\psi$ and $\tb_{\alpha\beta}$, implying $C^m$
differentiability of $\tH_{\alpha\beta}$. The idea of the proof is simple.
Equations (\ref{BC9}, \ref{BC10}) are singular at $x=0$, but we can
use $C^{m+1}$ regularity to control the behaviour of the highest
derivatives on the right-hand sides. We can then eliminate the
highest derivatives in these equations by setting $x=0$.
The derivatives remaining on the right are then of lower
order than those on the left, so we can proceed by induction.

To begin, observe that every term on the right-hand
side of (\ref{BC10}) can be written in terms of $\tb_{\alpha\beta}$, its
inverse, its first $k+1$ derivatives, and $\psi$ and its first
$k+1$ derivatives. Thus, for $k\le m$ we can use $C^{m+1}$
regularity to set $x=0$ in (\ref{BC10}), thereby eliminating the
highest derivatives and obtaining
\begin{eqnarray}
\der^k\psi \bigg\vert _0
&=&\frac{(k-1)}{(n-1)}\der^{k-2} \left ( \psi^2{\der {\tH}}
-\psi^2 \left [ D^2 +\tH_{\alpha\beta}\tH^{\alpha\beta}\right ]\psi
\right )\bigg \vert_0\label{BC11}\\
&=:&F_{(k)} \left ( \tb_{\alpha\beta}(0),\tH_{\alpha\beta}(0),\dots,
\tH_{\alpha\beta}^{(k-1)}(0),\psi(0),\dots,\psi^{(k-2)}(0) \right )
\quad ,\label{BC12}
\end{eqnarray}
where the function $F_{(k)}$ is defined for $k\ge 2$, a superscript
$p$ in parentheses denotes the $x$-derivative of order $p$, and the
subscript $0$ denotes evaluation at $x=0$. Note that $F_{(k)}$ can
depend on {\it tangential} derivatives of its arguments (through
the $D^2\psi$ term), although our notation does not make that explicit.

We want only order $k-1$ derivatives in $\tb_{\alpha\beta}$ and $\psi$ in
$F_{(k)}$, but the appearance of $\tH_{\alpha\beta}^{(k-1)}$ prevents this.
We can, however, express $\tH_{\alpha\beta}^{(k-1)}$ in terms of lower order
derivatives by using equation (\ref{BC9}). Now every term in
(\ref{BC9}) can be written as a combination of $\tb_{\alpha\beta}(x)$,
its inverse, its derivatives, and $\psi$ and its derivatives,
with the highest order derivatives appearing on the right-hand
side, both of $\tb_{\alpha\beta}$ and of $\psi$, being of order $k+1$.
Therefore, provided $k\le m$ and $k\neq n$, we can take $x=0$
in (\ref{BC9}), again eliminating the highest derivatives. We obtain
\begin{eqnarray}
\partial_x^{k-1}\tH_{\alpha\beta}\bigg \vert_0
&&=\frac{(k-1)}{(n-k)}\,\, \partial_x^{k-2} \left ( \left [
2\tH_\alpha {}^{\gamma}\tH_{\beta\gamma}-\tH\tH_{\alpha\beta}
+{\cal R}_{\alpha\beta}-D_\alpha D_\beta\right ]\psi \right )
\bigg \vert_0 \qquad \qquad \label{BC13}\\
&&=:G_{(k)\alpha \beta}\bigg
(\tb_{\alpha\beta}(0),\tH_{\alpha\beta}(0),
\dots,\tH_{\alpha\beta}^{(k-2)}(0),\psi(0),\dots, \psi^{(k-2)}(0)
\bigg )\ ,\qquad \qquad\label{BC14}
\end{eqnarray}
where the function $G_{(k)\alpha \beta}$ is defined for $k\ge 2$
and $k\neq n$. Thus, beginning at $k=2$ and excepting $k=n$, the
system comprised of (\ref{BC12}) and (\ref{BC14}) expresses the
order $k$ $x$-derivatives of $\psi$ and $\tb_{\alpha\beta}$ (order
$k-1$ $x$-derivatives of $\tH_{\alpha\beta}$) at $x=0$ in terms of
the lower order $x$-derivatives at $x=0$.

We may solve these equations iteratively, beginning with $k=2$. To
start the iteration, we must supply the data $\psi(0)$, ${\der
\psi}(0)$, $\tb_{\alpha\beta}(0)$, and ${\der {\tb}}_{\alpha
b}(0)$ or equivalently $\tH_{\alpha\beta}(0)$. However, equation
(\ref{BC8}) fixes $\tH_{\alpha\beta}(0)={\der \psi}(0)=0$, while
$\psi(0)$ is determined by (\ref{FE7}) and the definition
$\tW=1/\psi^2$ to be
\begin{equation}
\psi(0)=\sqrt{\frac{n(n-1)}{-2\Lambda}}\equiv\ell\quad .\label{BC15}
\end{equation}
The iteration proceeds until $k=n$, at which point it fails to
assign a value to $\tH^{(n-1)}_{\alpha\beta}(0)$ or equivalently to
$\tb_{\alpha\beta}^{(n)}(0)$. If a value for this quantity is assigned by
fiat, the iteration can again proceed until the limit imposed by
the assumed differentiability class of the Fermat metric is
reached. Thus, the free data are $\tb_{\alpha\beta}(0)$ and, if $k\ge n$,
$b^{(n)}_{\alpha\beta}(0)$ (equivalently, $\tH^{(n-1)}_{\alpha\beta}(0)$)
as well.

Finally, since $\psi(0)=\ell$ is constant on the $x=0$ surface and
since we assume ${\cal R}_{\alpha\beta}(0)=0$, then one can see by
applying the Leibniz rule in the definitions of $F_{(k)}$ and
$G_{(k)\alpha \beta}$ that
\begin{equation}
F_{(k)}(\tb_{\alpha\beta}(0),0,\dots,0,\psi(0),0,\dots,0)
=G_{(k)\alpha\beta}(\tb_{\alpha\beta}(0),0,
\dots,0,\psi(0),0,\dots,0) =0\quad .\label{BC16}
\end{equation}
Thus, if all derivatives of $\tb_{\alpha\beta}$ and $\psi$ below order $k$
vanish at $x=0$, then so do the order $k$ derivatives, unless
$k=n$. \qed

\noindent {\bf Corollary II.2.2.} {\sl If $(\Sigma,h,N)$ is $C^{n+1}$
conformally compactifiable with Ricci flat conformal boundary $x=0$,
where $n:={\rm dim}\Sigma\ge 3$, then the Fermat scalar curvature $\tR$
and its first $n-3$ $x$-derivatives vanish at $x=0$.}

\noindent {\bf Proof.} For $1<k<n$, the definition $\psi=1/\sqrt{\tW}$
yields
\begin{equation}
\der^k\tW = -2\frac{\der^k\psi}{\psi^3}+\dots+(-1)^{k}(k+1)!
\frac{(\der\psi)^k}{\psi^{k+2}}\quad ,\label{BC17}
\end{equation}
where the dots represent $k-2$ terms, all containing $x$-derivatives
of $\psi$ of order $<k$. Then Proposition II.2.1 implies the
vanishing of the right-hand side of (\ref{BC17}) at $x=0$. But,
differentiating (\ref{FE7}) $k\le n$ times and setting $x=0$, we
obtain
\begin{equation}
\der^k\tW\bigg \vert_0
=\frac{-k(k-1)}{n(n-1)}\tR^{(k-2)}(0)\quad ,\label{BC18}
\end{equation}
for $k\ge 2$. Since we have just shown that the left-hand side of
this expression vanishes for $k<n$, so does the right-hand
side, and so the first $n-3$ $x$-derivatives of $\tR$ vanish at $x=0$
as claimed.\qed

{}From (\ref{BC18}), (\ref{BC17}), and (\ref{BC12}), we obtain
\begin{equation}
\tR^{(n-2)}(0)=\frac{2}{\psi^3(0)}F_n\left (
\tb_{\alpha\beta}(0),0,\dots,0,\tH^{(n-1)}_{\alpha\beta}(0),\psi(0),0,
\dots,0\right ) \quad .\label{BC19}
\end{equation}
Thus, $\tR^{(n-2)}(0)$ and $\tH^{(n-1)}_{\alpha\beta}(0)$ are related,
so we may regard $\tR^{(n-2)}(0)$ as part of the free data. We will soon
see that $\tR^{(n-2)}(0)$ encodes the mass, but first we determine
the exact relation between it and $\tH^{(n-1)}_{\alpha\beta}(0)$:

\noindent {\bf Corollary II.2.3.} {\sl If $(\Sigma,h,N)$ is $C^{n+1}$
conformally compactifiable with Ricci flat conformal boundary $x=0$,
then the Fermat mean curvature of constant lapse surfaces near $x=0$
obeys}
\begin{equation}
\tH(x)=\frac{\ell}{2}\frac{x^{n-1}}{(n-1)!}\tR^{(n-2)}(0)+{\cal O}(x^n)
\quad . \label{BC20}
\end{equation}

\noindent{\bf Proof.} If we contract (\ref{BC4}) with
$\tb^{\alpha\beta}$, we obtain
\begin{equation}
\tR_{ab}\tn^a\tn^b-\tR=-(n-1)\frac{\tH}{x\psi}\quad .\label{BC21}
\end{equation}
If we use (\ref{FE11}), (\ref{BC21}), and the Gauss formula
\begin{equation}
\tR_{ab}\tn^a\tn^b - \frac{1}{2}\tR =\frac{1}{2}\left [ \tH^2
-\tH^{\alpha\beta}\tH_{\alpha\beta}-{\cal R}\right ]\quad
,\label{BC22}
\end{equation}
where ${\cal R}$ is the scalar curvature of the slice $V_x$ in the
induced metric, then we can eliminate $\tR_{ab}\tn^a\tn^b$ and
$\tR$ to obtain
\begin{equation}
\tH=\frac{x\psi}{(n-2)}\left [ \frac{x}{2(n-1)} \frac{\partial\tR}
{\partial x}+{\cal R}-\tH^2+\tH_{\alpha\beta}\tH^{\alpha\beta} \right ]
\quad .\label{BC23}
\end{equation}
{}From point ({\it ii}) of Proposition II.2.1 (and using ${\cal R}(0)=0$),
we can write that $\tH_{\alpha\beta}={\cal O}(x^{n-1})$,
$\psi=\ell+{\cal O}(x^n)$, and ${\cal R}={\cal O}(x^n)$, so
\begin{equation}
\tH=\frac{x^2\ell}{2(n-1)(n-2)}\frac{\partial\tR}{\partial x}
+{\cal O}(x^{n+1}) \quad .\label{BC24}
\end{equation}
Finally, Corollary II.2.2 implies that
\begin{equation}
\frac{\partial \tR}{\partial x}=\frac{x^{n-3}}{(n-3)!} \tR^{(n-2)}(0)
+{\cal O}(x^{n-2})\quad .\label{BC25}
\end{equation}
Substitution of (\ref{BC25}) into (\ref{BC24}) yields (\ref{BC20}).\qed

\subsection{The Mass}
\setcounter{equation}{0}

\noindent We now show that $\tR^{(n-2)}(0)$ is, up to a factor,
the ``mass aspect'' whose integral over the conformal boundary
agrees with the Ashtekar-Magnon conformal definition of mass.
\footnote {In \cite{dHSS}, $\tb^{(n)}_{\alpha\beta}$ was related
to a new definition of mass.}
In Appendix A, we show that it is similarly related to the
Hawking-Horowitz mass. Equivalence to various other AdS mass
definitions then follows \cite{HH}. In view of Corollary II.2.3,
this establishes that the sign of the mass aspect governs the mean
convexity/concavity of constant $x$ surfaces near infinity.

For any spacetime of dimension $n+1\ge 4$ with metric $g_{ab}$ and
Riemann tensor $P^a{}_{bcd}$, the Weyl tensor $C_{abcd}$ is defined by
\begin{eqnarray}
C_{abcd}&:=& P_{abcd}-\frac{1}{n-1} \left ( g_{bc}S_{ad}
-g_{bd}S_{ac}+g_{ad}S_{bc}-g_{ac}S_{bd} \right )\label{BC26} \quad ,\\
S_{ab}&:=&P_{ab}-\frac{1}{2n}g_{ab}P\label{BC27}\quad .
\end{eqnarray}
Under the conformal transformation $\tg_{ab}=\Omega^2 g_{ab}$,
$S_{ab}$ obeys
\begin{equation}
S_{ab}=\tS_{ab}+\frac{(n-1)}{\Omega}\tnabla_a\tnabla_b\Omega
-\frac{(n-1)}{2\Omega^2}\tg_{ab}
\tg^{cd}\tnabla_c\Omega\tnabla_d\Omega\quad .\label{BC28}
\end{equation}
We are interested in the particular case where the metric is the
conformally rescaled spacetime metric $\tg_{ab}:=x^2 g_{ab}$,
with $g_{ab}$ as in equation (\ref{FE1}),
so $\Omega=1/N=:x$. We maintain our convention of having tildes
denote quantities defined with respect to a rescaled metric. We
observe that $\tg_{ab}$ has a unit timelike Killing field, which
we denote $\ttime^a$, $\tg_{ab}\ttime^a\ttime^b=-1$, and therefore
$\tP_{abcd} t^at^c=0$. We now contract (\ref{BC26}) with $\ttime^a
\tn^b \ttime^c \tn^d$, where $\tn^a$ is the unit (in $\tg_{ab}$)
normal to the constant $x$ surfaces (thus $\tg_{ab}\tn^a\tn^b=+1$,
$\tg_{ab}\ttime^a\tn^b=0$), obtaining
\begin{equation}
\frac{1}{n-1}\left ( \tS_{ab} \tn^a \tn^b -\tS_{ab} \ttime^a \ttime^b
\right )
=\tWeyl_{abcd}\ttime^a \tn^b \ttime^c \tn^d
=\frac{1}{x^2}C_{abcd} t^a n^b t^c n^d
=\frac{1}{x}E_{ac}t^at^c\label{BC29}\quad ,
\end{equation}
where $C_{abcd}$ and $t^a=x\ttime^a$ are, respectively, the Weyl
tensor and timelike Killing field of the unrescaled spacetime
metric, and $E_{ab}$ is the electric part of $C_{abcd}$ with
respect to $n^a = x\tn^a$.\footnote
{The Ashtekar and Magnon \cite{AM} definition of $E_{ab}$ differs
from ours by a factor of $\ell^2$. This compensates for the fact
that they use a normal vector $n^a$ of magnitude $1/\ell$, whereas
we use a unit normal.}

To evaluate the left-hand side, we use the condition that the
spacetime metric $g_{ab}$ is Einstein, which here we can write as
$S_{ab}= (\Lambda/n) g_{ab}$. Then (\ref{BC28}) becomes
\begin{eqnarray}
\tS_{ab}&=& -\frac{(n-1)}{x}\tnabla_a\tnabla_b x + \frac{1}{2nx^2}
\tg_{ab}\left [ n(n-1)\tW+2\Lambda\right ]\nonumber\\
&=&\tR_{ab}-\frac{1}{2n}\tR\tg_{ab}\quad ,\label{BC30}
\end{eqnarray}
where in the last step we used equations (\ref{FE4}) and
(\ref{FE7}). Then it is easy to evaluate the
left-hand side of (\ref{BC29}), yielding
\begin{equation}
\frac{1}{x}E_{ac}t^at^c = \frac{1}{n-1}\left ( \tR_{ab}
\tn^a \tn^b - \frac{1}{n}\tR \right )\quad . \label{BC31}
\end{equation}
But we can use the identity (\ref{FE11}) to rewrite the right-hand
side, obtaining
\begin{eqnarray}
E_{ac}t^at^c &=&\frac{x^2}{2n(n-1)} \frac{\partial
\tR}{\partial x} =\frac{x^2}{2n(n-1)}\left [ \frac{x^{n-3}}{(n-3)!}
\tR^{(n-2)}(0) + {\cal O}(x^{n-2})
\right ] \nonumber\\
&=&\frac{(n-2)}{2(n!)}x^{n-1}\tR^{(n-2)}(0) + {\cal O}(x^n)
\quad ,\label{BC32}
\end{eqnarray}
where we used Corollary (II.2.2) and assumed that the metric is $C^{n+1}$.
We are now in a position to prove the following:

\noindent {\bf Proposition II.3.1.} {\sl If $(\Sigma,h,N)$ obeys
equations (\ref{FE2}--\ref{FE3}) and is $C^{n+1}$ conformally
compactifiable, then the generalized (in dimension $n$)
Ashtekar-Magnon mass $M_{AM}$ of $\Sigma$ is given by}
\begin{equation}
M_{AM}=\frac{-\ell}{16\pi n!} \int_{\partial \tSigma}
\frac{\partial^{n-2} \tR}{\partial x^{n-2}}\bigg|_0\sqrt{\tb}\ dS
=\frac{-\ell{\rm vol} (\partial \tSigma)}{16\pi n!}
\left \langle \frac{\partial^{n-2} \tR}{\partial x^{n-2}}\bigg|_0
\right \rangle \quad ,\label{BC33}
\end{equation}
{\sl where ${\rm vol}(\partial \tSigma)=\int_{\partial \tSigma}\sqrt{\tb}\ dS$
and angle brackets denote the average over $\partial \tSigma$ with respect
to the measure $\sqrt{\tb}\ dS$.}

\noindent{\bf Proof.} Let $t^a$ be a timelike Killing vector field
of the spacetime metric. Let $\Sigma$, $\partial \tSigma$ be as above. Then
we define the Ashtekar-Magnon mass (cf.\ \cite{AM}, eq.\ (11) for
the $n=3$ case) by
\begin{equation}
M_{AM}:=\frac{-\ell}{8\pi (n-2)} \int_{\partial \tSigma} E_{ab} t^a t^b
\sqrt{b}\ dS \quad ,\label{BC34}
\end{equation}
where $\sqrt{b}\ dS=\sqrt{\tb}\ dS/x^{n-1}$ is the measure induced
on $\partial \tSigma$ by the (unrescaled) metric on a constant $x$
surface (so $\sqrt{\tb}\ dS$ is the measure induced by the
Fermat metric) and the limit $x\to 0$ is to be taken. Since we
assume the metric is $C^n$, we may use (\ref{BC32}), from which we
obtain
\begin{equation}
M_{AM}=\frac{-\ell}{16\pi n!} \int_{\partial \tSigma}
\tR^{(n-2)}(0)\sqrt{\tb}\ dS
\quad ,\label{BC35}
\end{equation}
in the limit as $x\to 0$.\qed

\noindent {\bf Definition II.3.2.} In light of Proposition II.3.1,
we define the {\it mass aspect} of a static, negative mass
spacetime with Ricci flat conformal boundary at $x=0$ to be
\begin{equation}
\mu:=\frac{-\ell}{2(n!)}\tR^{(n-2)}(0)\quad .\label{BC36}
\end{equation}

\noindent Then notice from Corollary II.2.3 that negative mass aspect
implies that surfaces $x=\epsilon=const $ are mean (outward) convex
for small enough $\epsilon>0$.

\section{A Geometric Structure Theorem and Negative Mass}
\label{structure.sec}
\setcounter{equation}{0}

\noindent En route to the uniqueness theorem for the soliton, we will
obtain a more general structure result for static spacetimes that obey
a convexity condition on the extrinsic geometry of constant lapse
surfaces near scri. Section III.1 discusses the convexity condition
and relates it to negative mass, while III.2 contains the
structure theorem and its proof.

\subsection{Convex Surfaces of Constant Lapse} \label{cc}

\noindent Consider the level surfaces of the lapse $N$
(equivalently, of $x:=\tN:=1/N$). Recall that the second
fundamental form $\tH_{\a\b}$ of each level surface is defined in
Section II using the Fermat ``outward'' unit normal vector
pointing towards scri. The eigenvalues of $\tH_{\a\b}$ are called
the \emph{principal curvatures}.

\noindent {\bf Definition III.1.1.} We say that $(\Sigma,h,N)$
\emph{satisfies condition (S)} provided that
the second fundamental form $\tH_{\a\b}$ of each level surface $N=c$
is semi-definite (equivalently, provided that the principal curvatures
of each level surface $N=c$ are either
all non-negative or all nonpositive) whenever $c$ is sufficiently
large (i.e., near scri). If $\tH_{\a\b}$ is positive
semi-definite (equivalently, if the principal curvatures are all
non-negative) for each of the level surfaces in this neighbourhood of scri,
we say that $(\Sigma,h,N)$
\emph{satisfies condition (C)}, and the level surfaces of $N$ in this
neighbourhood are said to be {\it weakly convex}.

In the next subsection, we will use Condition (C) to control the
behaviour of certain geodesics near scri as follows.
Suppose Condition (C) holds, so that the level surfaces $N=c$ are weakly
convex, in the sense of the definition, for all $c$ sufficiently large.
Let $V= \{N=c_0\}$ be such a level surface; $V$ has a well defined ``inside''
($N<c_0$) and ``outside'' ($N>c_0$). Then, as follows from the maximum
principle, if $\gamma$ is a geodesic segment with endpoints inside $V$,
all of $\gamma$ must be contained inside $V$.
Thus, Condition~(C) provides ``barrier surfaces'' for the construction
of certain minimizing geodesics, as will be seen in the next subsection.

Condition (S) allows the level surfaces $N = c$ near scri to be
either mean convex ($H_{\a\b}$ positive
semi-definite) or mean concave ($H_{\a\b}$ negative semi-definite).
All the relevant examples known to us obey condition (S), even
when Condition (C) fails. As the following lemma indicates, when Condition
(S) holds, the sign of the mass aspect $\mu$ determines, in the case of
interest here, whether one gets weakly convex or weakly concave surfaces.

\noindent {\bf Lemma III.1.2.} {\sl If $\mu<0$ pointwise on the Ricci
flat conformal boundary at $x=0$ of a $C^{n+1}$ conformally
compactifiable spacetime and if Condition (S) holds, then Condition
(C) holds.}

\noindent {\bf Proof.} From Corollary II.2.3 and Definition III.1.1,
we have that $\mu<0$ implies $\tH(1/c)>0$ whenever $c>C$ for some
$C\in\re$, so the sum of the principal curvatures of each $x=1/c$
level set is positive. By Condition (S), the principal curvatures
are either all nonnegative or all nonpositive, so that sign is
nonnegative, implying Condition (C).\qed

\subsection{The Structure Theorem}

\noindent {\bf Theorem III.2.1.} {\sl Consider an
$(n+1)$-dimensional static spacetime as in (\ref{FE1}) such that
({\it i}) $(\Sigma,h,N)$ is smoothly ($C^{\infty}$) conformally compactifiable,
({\it ii}) the static vacuum field equations hold, and ({\it iii}) condition
(C) holds. Then the Riemannian universal cover $({\tSi }^\ast,
\wh^\ast)$ of the conformally related spacetime $(\tSi,\wh)$
splits isometrically as
\begin{equation}
\tSi^\ast = \mathbb R^k\times \tW, \qquad {\wh}^\ast = h_E\oplus
\tsigma\,\label{ST1}
\end{equation}
where $(\mathbb R^k, h_E)$ is standard $k$-dimensional Euclidean
space, with $0\leq k \leq n$, and $(\tW,\tsigma)$ is a compact
Riemannian manifold with non-empty boundary. The Riemannian
universal cover $(\Sigma^\ast , h^\ast)$ of $(\Sigma,h)$ splits
isometrically as a warped product of the form
\begin{equation}
\Sigma^\ast = \mathbb R^k\times W, \qquad h^\ast = ({N^\ast}^2
h_E)\oplus \sigma \, ,\label{ST2}
\end{equation}
where $N^\ast=N\circ \pi$ ($\pi =$ covering map) is constant on $\re^k$,
and $(W,\sigma)$ is a simply connected Riemannian manifold such
that $(W,\sigma,N)$ is smoothly conformally compactifiable. }

\noindent {\bf Remark III.2.2.} Notice that we specify no boundary data in
this theorem, instead imposing only the convexity condition (C). In particular,
this theorem does not require Ricci flatness of the conformal boundary.
However, if the conformal boundary is Ricci flat and if the mass is
negative, we may relax Condition (C) to Condition (S), obtaining in
this case a structure theorem for negative mass static spacetimes.

\noindent {\bf Remark III.2.3.}
Theorem III.2.1 is similar in spirit to the Cheeger-Gromoll splitting
theorem \cite{CG}, or more precisely, to Theorem 3.16 in \cite{PP},
a structure theorem for compact (without boundary) Riemannian manifolds
of nonnegative Ricci curvature, which is a direct consequence of the
Cheeger-Gromoll splitting theorem. Theorem III.2.1 implies a
strong structure result for the fundamental group $\pi_1(\tSi)$ of $\tSi$.
For example, if $k=0$ then the universal cover is compact, and $\pi_1(\tSi)$
is finite. More generally, Theorem III.2.1 implies that $\pi_1(\tilde \Sigma)$
is ``almost abelian'', i.e., that there exists a finite
normal subgroup $F$ of $\pi_1(\tilde \Sigma)$ such that
$\pi_1(\tilde \Sigma)/F$ contains a subgroup of
finite index isomorphic to $\mathbb Z^k$, cf.\ \cite{CG}.

\noindent {\bf Proof.} The proof of this theorem consists of three
parts. We first show that a null line exists in the universal
covering spacetime whenever $\Sigma^\ast$ is noncompact. The null
splitting theorem then tells us that this null line lies in a
smooth, closed, achronal, totally geodesic null hypersurface
$\cH$. Staticity then implies that $\cW_t =\Sigma_t^\ast \cap \cH$
is totally geodesic, where $\Sigma_t^\ast$ is a constant time
slice. The $t=0$ slice $\Sigma_0^\ast$ can thus be foliated by the
projections of the $\cW_t$ into it. Using (\ref{FE2}, \ref{FE3}),
we show that we can then isometrically split off an $\re$
factor, and continuing iteratively yields the result.

%\vfill\par\eject

\noindent ({\it i}) \underbar {Construction of a null line:}

\noindent
A \emph{line} in a Riemannian manifold is a complete geodesic, each segment
of which is minimal (length minimizing). We describe here how to construct
a line in $(\tSi^\ast,\wh^\ast)$, provided $\tSi^\ast$ is noncompact. We
then make use of a fundamental feature of the Fermat metric: Length
minimizing Fermat geodesics lift in an essentially unique way to achronal
null geodesics in the physical spacetime, see Appendix B.

Let $(\tSi^\ast,\wh^\ast)$ be the Riemannian universal
cover of $(\tSi,\wh)$. If $\tSi^\ast$ is compact, then $k=0$ in
the above splitting and we are done, so assume otherwise. Let $p
\in {\rm int}(\tSi^\ast)$ and let $\{q_i\}$ be a sequence of
points bounded away from $\pd \tSi^\ast$, such that the
distances from $p$ to successive $q_i$ tend to infinity. For each
$i$, let $\gamma_i$ be a minimal geodesic
from $p$ to $q_i$. The convexity condition (C) implies that the
$\gamma_i$'s are uniformly bounded away from $\pd \tSi^\ast$. Fix a
fundamental domain $D\subseteq\Sigma^\ast$. For each midpoint
$r_i$ of $\gamma_i$, there is a covering space transformation
$\fg_i \in\pi_1(\tSi)$, possibly the identity, mapping $r_i$ into
$D$. Because the $\fg_i$ are discrete isometries, the images
${\hat \gamma}_i=\gamma_i\circ\fg_i$ form a sequence of
minimal geodesics that all meet $D$ and remain uniformly bounded away from
$\pd \tSi^\ast$. Since $\widetilde \Sigma$ is compact then so is
$D$, and so there will be a convergent subsequence of the ${\hat
\gamma}_i$ whose limit is a complete, minimal geodesic (thus a
line) $\gamma$ of the Fermat metric $\wh^*$, meeting $D$,
which is bounded away from $\pd \tSi^\ast$. Finally,
fix a point on $\gamma$. By Lemma B.1 in Appendix B, through that point
there is a unique, future directed null line $\eta$ in
$(\Sigma^\ast, h^\ast, N^\ast)$ produced by lifting $\gamma$ along
the timelike Killing field.

\noindent ({\it ii}) \underbar{Splitting off one $\mathbb R$ factor:}

Let $\lambda\to\mu(\lambda)$, $\lambda\in I$, be an inextendible
null geodesic in the physical spacetime $({\cal M},g)$. Using the
constant of motion lemma with respect to the Killing field
$\pd/\pd t$ and the fact that $\mu$ is null, one obtains along
$\mu$ that $d\lambda = Nds$ (up to a factor constant along $\mu$)
and $s$ is $h$-arclength along the projection of $\mu$ into
$\Sigma$. Note that since $(\Sigma,h,N)$ is conformally
compactifiable, $(\Sigma,h)$ is necessarily geodesically complete
and $N$ is bounded positively away from zero. From the geodesic
completeness of $(\Sigma, h)$ and the inextendibility of $\mu$, it
follows that $s\to \pm\infty$ as $\lambda$ ranges over $I$. Since
$N$ is bounded away from zero, the equation $d\lambda = Nds$ then
implies that $\lambda$ ranges over all real numbers, i.e., $I
=\re$. Thus, $({\cal M},g)$ is null geodesically complete, and
this completeness lifts to the universal covering spacetime.

Because the null geodesically complete covering spacetime
$(\Sigma^\ast, h^\ast, N^\ast)$ satisfies the null energy
condition and admits the null line $\eta$, the null splitting
theorem I.1 implies that $\eta$ is contained in a smooth,
connected, achronal, edgeless totally geodesic null surface $\cH$.
By the constant of motion lemma, the equation $dt/d\lambda =
E/(N^\ast)^2$ holds along $\lambda \mapsto\eta(\lambda)$, where
$\lambda$ is an affine parameter, $E$ is constant along $\eta$,
and, because $\gamma$ is bounded away from $\pd \tSi^*$, $N^*$ is
bounded along $\eta$. Since $\eta$ is complete it follows that $t$
ranges over all real numbers. Thus $\eta$ meets each constant $t$
hypersurface $\Sigma_t^*= \{t\} \times \Sigma^*$. Since each
$\Sigma_t^*$ is totally geodesic, the codimension 2 spacelike
intersections $\cW_t=\Sigma^\ast_t \cap \cH$ are also totally
geodesic.

Let $\mu$ be any other null generator of $\cal H$ passing through ${\cal W}_0$.
Since $\cal H$ is totally geodesic,
its null generators have zero expansion and shear. It follows that the
spatial separation of $\mu$ and $\eta$
remains constant along these generators. Hence, the projection of $\mu$
into $\Sigma^*$ will also be
bounded away from $\pd \tSi^*$, and by the same
argument as above, $t$ will range over all reals along $\mu$. Thus,
the null generators of ${\cal H}$ meeting ${\cal W}_0$ can be parametrized
by time $t\in\re$ so as to define a flow ${\cal F}:\re\times \cW_0\to{\cal H}$
along $\cH$. Using that $\cal H$ is connected and closed, it can be shown
that ${\cal F}$ is onto. Hence, fixing any $t\in\re$, the flow induces a
diffeomorphism $\cW_0\to \cW_t$.

Let ${\cal P}$ denote projection into $\Sigma_0^\ast$ along
integral curves of $\partial/\partial t$ and let $W_t:={\cal
P}(\cW_t)$ (note that $W_0\equiv \cW_0$). The composition of
${\cal F}$ with ${\cal P}$ defines a flow $F:={\cal P}\circ {\cal
F}: \re\times W_0\to\Sigma_0^\ast$ on the $t=0$ hypersurface. The
$t$ parametrization on ${\cal F}$ gives a parametrization for $F$
by arclength in the Fermat metric $\wh^\ast:= h^\ast/ (N^\ast)^2
=({\cal P}g)/(N^\ast)^2$. The equipotentials are ${\cal
P}(\cW_t)\equiv W_t$ and are all diffeomorphic copies of $W_0$.
The flow lines $t \mapsto F(t,q)$, $q\in W_0$, being projections
of the null generators of $\cal H$, are Fermat geodesics
orthogonal to the $W_t$s. We prefer to relabel the parameter along
the flow $F$ as $u$ from here onward, so we write the
equipotentials as $W_u$ and the flow field orthogonal to them as
$\partial/\partial u$. Since ${\cal P}|_{{\cal H}}$ is both an
open and closed mapping, it follows that ${\cal P}({\cal H}) =
\Sigma_0^*$. Thus, $F$ maps $\re\times W_0$ diffeomorphically onto
$\Sigma_0^*$, i.e., $\Sigma_0^*\approx \re\times W_0$. Finally, we
pull this back along the embedding $i:\Sigma^\ast\to
\Sigma_0^\ast$ to obtain $\Sigma^\ast \approx \re \times W$ where
$W_0=i(W)$. Thus, we have extracted the first topological factor
of $\re$. We now turn our attention to the geometrical splitting.

In coordinates adapted to the flow, which in fact are just
Gaussian normal coordinates about $W_0$, the Fermat metric takes the form
\beq\label{gauss}
\wh^\ast = du^2 +\tsigma_{ij}dy^i dy^j\label{ST3} \quad,
\eeq
where $u\in\re$, $\vec y$ represents
coordinates on $W_0$, and $\tsigma_{ij}(u, \vec y)$ is the induced
metric on $W_0$. The physical metric $h^\ast$ is then
\begin{equation}
h^\ast= N^{\ast 2}(u,{\vec y})du^2 + \sigma_{ij}(\vec y)dy^i
dy^j\quad , \label{ST4}
\end{equation}
where, since each $W_u$ is totally geodesic in the physical metric, the
induced metric $\sigma_{ij}(\vec y)=N^{\ast 2} \tsigma_{ij}$ on
$W_0$ is $u$-independent.

We now show that the lapse $N^\ast$ is independent of $u$. Let
$N^a=\frac{1}{N^{\ast}} (\frac{\pd}{\pd u})^a$ be the unit normal
to the surfaces $W_u$. Since the $W_u$ are totally geodesic, the
Codazzi relation yields $R^\ast_{ab}N^ah^{\ast b}_c=0$, so that
$R^\ast_{ab}N^aY^b=0$ for any $Y^a$ tangent to $W_u$. Using this
in eq. (\ref{FE2}), we get
\begin{equation}
N^aY^b\nabla_a\nabla_b N^\ast =0 \quad ,\quad \forall Y^a\in TW_u
\label{ST5}\quad .
\end{equation}
But the surfaces $W_u$ are totally geodesic, so the Hessian in
(\ref{ST5}) becomes the double directional derivative $\nabla_N
(\nabla_Y N^\ast)$. Since this vanishes for all $Y^a\in TW_u$, the
general solution is $N^\ast=\alpha(u)\beta(\vec y)$.
Contracting equation (\ref{FE2}) with $N^a N^b$ yields,

%\newpage
\begin{eqnarray}
R_{ab}^\ast N^aN^b &=& \frac{1}{\alpha^2\beta^2}R_{uu}^\ast\nonumber\\
&=&\frac{1}{\alpha^2\beta^2}\biggl(\frac{1}{\alpha}\frac{d^2\alpha}{du^2}
-\frac{1}{\alpha^2}\left ( \frac{d\alpha}{du}\right )^2 +\alpha^2
\sigma^{ij}\frac{\partial \beta}{\partial y^i} \frac{\partial
\beta}{\partial y^j} + \frac{2\Lambda}{(n-1)}
\alpha^2\beta^2\biggr)\ .\qquad\qquad\label{ST6}
\end{eqnarray}
Since $N=\alpha(u)\beta({\vec y})$, we can make the coordinate
transformation $v= \int \alpha du$,
and since $\alpha>0$ is uniformly bounded above and away from zero
below and $u$ takes values throughout $\mathbb R$, so does $v$.
Notice that $(\frac{\pd}{\pd v})^a$ is a Killing vector in $h^\ast$
(but not necessarily in the full spacetime). In the new coordinates
$(v,\vec y)$, we have $R_{uu}^\ast=\alpha^2R_{vv}^\ast$, where
$R_{vv}^\ast$ is independent of $v$. Eq.\ (\ref{ST6})
is thus separable in $v$ and $\vec y$, and takes the form
\begin{equation}
R_{vv}^\ast -{\sigma}^{ij}\frac {\partial \beta}{\partial y^i}
\frac{\partial\beta}{\partial y^j} - \frac{2\Lambda}{(n-1)}
\beta^2 = \frac{1}{\alpha^3}\frac{d^2\alpha}{du^2}
-\frac{1}{\alpha^4}\left ( \frac{d\alpha}{du}\right )^2
=\frac{1}{\alpha(v)}\frac{d^2\alpha}{dv^2}\quad ,\label{ST7}
\end{equation}
yielding in particular
\begin{equation}
\frac{1}{\alpha(v)}\frac{d^2\alpha}{dv^2}=c\quad ,\label{ST8}
\end{equation}
where $c$ is the separation constant. Obviously the only solution
suitably bounded on all of $\mathbb R$ is $\alpha=const.$, which
occurs only for $c=0$. This in turn implies that the lapse $N^\ast$
is independent of $u$, which implies that the $W_u$ are totally
geodesic in the Fermat metric.

The spacetime $(\Sigma^\ast, h^\ast,N^\ast)$ therefore splits as
$\Sigma^\ast = \re \times W$, and $h^\ast={N^\ast}^2 du^2 \oplus
\sigma$, where $N^\ast$ and $\sigma$ are $u$-independent.
$(\tSi^\ast, \wh^\ast, \tN^\ast)$ then splits as a product
$\tSi^\ast = \re \times \tW$, $\wh^\ast= du^2 \oplus\tsigma$
(the splitting clearly extending to the boundary).

\newpage

\noindent ({\it iii}) \underbar {Iteration:}

\noindent If $\tW$ is compact then we are done and $k=1$. If not,
we proceed inductively. Assume that we have split off $p$ factors
of $\re$, so that
\begin{equation}
\tSi^\ast = \re^p \times\tW\quad ,\quad \wh^\ast=
\left ( \sum_{j=1}^p du_j^2\right )
\oplus \tsigma_{(n-p)}\quad ,\quad \frac{\partial
N^\ast}{\partial u_j} =0\quad ,\label{ST9}
\end{equation}
and $\tW$ is noncompact. We then
proceed to split off another factor of $\re$ as follows.

Let $\tW_{(n-p)} =\bigcap\limits_{j=1}^p \{u_j
=0\}\cap\tSigma_0^\ast$. We first show that since $\tW_{(n-p)}$ is
noncompact, not all the lines in $\tSi^\ast_0$ can lie in
$\re^p$. Let us assume otherwise, i.e., that all lines in
$\tSi^\ast_0$ have the form $\gamma(s) =(\gamma_1(s), x)$, where
$\gamma_1(s)$ is a line in $\re^p$ and $x \in \tW_{(n-p)}$. Under
the covering space isometry, $\fg\in \pi_1(\tSi)$, lines get
mapped to lines. Now, each vector in the tangent space of $\re^p$
is tangent to a line in $\re^p$. By assumption, all the lines lie
in the $\re^p$ factor, so that the endomorphism $\varphi$ on
$T\tSi_0^\ast$ induced by $\fg$ preserves the tangent space
$T\re^p \subset T\tSi_0^\ast$. Since $T\tW_{(n-p)}$ is orthogonal
to $T\re^p$ and $\varphi$ is linear, $\varphi$ preserves
$T\tW_{(n-p)}$ as well.
Since $\tW_{(n-p)}$ is noncompact,
we may repeat the construction, as in part ({\it i}), of a
sequence of minimal geodesic segments $\gamma_i$ in $\tW_{(n-p)}$, whose
lengths
diverge to infinity, and which are uniformly bounded away from
$\pd\tW_{(n-p)}$. Again, we use the covering transformations
$\fg_i$ to map the midpoints $r_i$ of $\gamma_i$, to the
fundamental domain $D$ of $\tSi_0^\ast$.
As before, the
minimal geodesics $\gamma_i'= \gamma_i\circ \fg_i$
meet the compact set $D$,
and, by passing to a subsequence if necessary, converge to a
line $\gamma$. By assumption, $\gamma$ lies in the factor $\re^p$.
Since $\fg_i$ maps the velocity vectors of $\gamma_i$ to the velocity
vectors of $\gamma_i'$, it follows that for large $i$, $\fg_i$ maps vectors
tangent
to $\tW_{(n-p)}$ to vectors nearly perpendicular to $\tW_{(n-p)}$.
This is a contradiction, so we conclude that not all lines lie in $\re^p$.

Since the $u_j$-directions are flat in the Fermat metric any line
$\gamma:(-\infty, \infty)\to \tSi_0^\ast = \re^p \times
\tW_{(n-p)}$ itself splits as $\gamma(s)=(\gamma_1(s),
\gamma_2(s))$, where $\gamma_1$ is a line in $\re^p$ and
$\gamma_2$ is a line in $\tW_{(n-p)}$---or possibly one of
$\gamma_1,\gamma_2$ is constant, but since not all lines lie in
the $\re^p$ factor, $(\tW_{(n-p)}, \tsigma_{(n-p)})$ itself must
contain at least one line. Call it $\tzeta$. Since the
hypersurfaces $u_j=0$ are all timelike,
$(W_{(n-p)},\sigma_{(n-p)}, N^\ast)$ is a static spacetime ${\cal
S}$ in its own right (even if it may not satisfy equations
(\ref{FE2}, \ref{FE3}), preventing us from applying the null
splitting theorem directly). Thus, $\tzeta$ can be lifted to a
null line $\zeta$ in ${\cal S}$.
Because the covering spacetime $(\Sigma^\ast,h^\ast, N^\ast)$
is the metric product $\re^p\times{\cal S}$,
$\zeta$ is also a null line in
$(\Sigma^\ast, h^\ast, N^\ast)$. The null splitting theorem can
now be applied to show that $\zeta$ lies in a smooth, edgeless,
achronal, totally geodesic null hypersurface $\cH_{p+1}$.

We now apply the construction of part ({\it ii}) to ${\cal
H}_{p+1}$ to produce a foliation of $\tW_{(n-p)}$. As in part
({\it ii}), there is a null geodesic flow on ${\cal H}_{p+1}$ that
projects to a flow on $\tSigma_0^\ast$. We must show that this
flow is tangent to $\tW_{(n-p)}$. Equivalently, the flow on ${\cal
H}_{p+1}$ must be tangent to ${\cal S}$. But this follows because
$\zeta\subset{\cal S}$, so ${\dot \zeta}^a\in T{\cal S}$, and the
flow vector field on ${\cal H}_{p+1}$ to which ${\dot \zeta}^a$
belongs is parallel (cf.\ Remark I.2(b)).

We thus obtain the splitting
$(\tW_{(n-p)},\tsigma_{(n-p)})=(\re\times \tW_{(n-p-1)},
du_{p+1}^2\oplus\tsigma_{(n-p-1)})$, with $\tsigma$ independent of
$u_{p+1}$, or $(W_{(n-p)}, \sigma_{(n-p)}) =(\re\times
W_{(n-p-1)}, {N^\ast}^2du_{p+1}^2 \oplus\sigma_{(n-p-1)})$.
Repeating the arguments of part ({\it ii}), we obtain that
$N^\ast$ is $u_{p+1}$-independent, and thus so is $\sigma$. Hence
we write
\begin{equation}
h^\ast = \left ( N^{\ast 2}\sum_{j=1}^{p+1} du_{j}^2\right )
\oplus \sigma \quad , \quad \frac{\pd N^\ast}{\pd u_j}=0\quad ,
\quad j=1, \dots, p+1\quad ,\label{ST10}
\end{equation}
where $\sigma$ is independent of the $u_j$. Since this procedure
is valid for any $p=1,\dots,n$, one can continue splitting off
$\re$ factors until what remains is compact.\qed

\section{The Uniqueness Theorem}\label{uniqueness.sec}
\renewcommand{\theequation}{\thesection.\arabic{equation}}
\setcounter{equation}{0}
{\bf Theorem IV.1.} {\sl
\label{uniqueness.th}
Consider a static spacetime as in (\ref{FE1}) such that $(\Sigma,h,N)$
{\rm (1)} is smoothly conformally
compactifiable, {\rm (2)} satisfies the static vacuum field equations, and
{\rm (3)} satisfies condition (S). In addition, assume the following
conditions to hold:
\begin{enumerate}
\item[\rm (a)] The boundary geometry of $(\tSi,\wh)$
is the same as that of (\ref{Intro1}), i.e., $\partial
\tSi = T^{n-2}\times S^1$, $\wh|_{\partial \tilde
\Sigma} = \sum\limits_{i=1}^{n-2}(d\theta^i)^2 + \frac1{\ell^2}d\phi^2$.
\item[\rm (b)] The mass aspect $\mu$ of $(\Sigma,h,N)$
is pointwise negative.
\item[\rm (c)]
Given the inclusion map $i:\partial\tSi \rightarrow \tSi$, the kernel of the
induced homomorphism of fundamental groups,
$i_*:\pi_1(\partial\tSi) \rightarrow \pi_1(\tSi)$, is
generated by the $S^1$ factor.
\end{enumerate}
Then the spacetime (\ref{FE1}) determined by $(\Sigma,h,N)$ is
isometric to the Horowitz-Myers soliton (\ref{Intro1}).
}

Assumptions (a) and (b) are natural boundary conditions for the
uniqueness problem. (Recall from Section II that the mass aspect
is related to the free boundary data.) As shown in \cite{CS}, in
$3+1$ dimensions, (b) automatically holds, provided $\mu$ is
constant. Assumption (c) pertains to a distinctive topological
feature of the AdS soliton. It asserts that the generator of the
$S^1$ factor in $(\pd\tSi,\wh|_{\partial \tilde \Sigma})$ is
contractible in $\tSi$, and moreover, that any loop in $\partial\tSi$
contractible in $\tSi$ is a multiple of the generator. As shown in
the proof, assumptions (a) and (c) together imply that $\pi_1(\tSi)
\approx \zz^{n-2}$. Were we to adopt the latter condition {\it en lieu}
of assumption (c), without assuming the geometrical $S^1$ factor is
contractible in the bulk, then one could still conclude via the
following proof that $(\Sigma,h,N)$ is \emph{locally} isometric to
the AdS soliton (the universal covers will be isometric).
For example, this situation can arise when some loop not in the
$\pd\tSi$ homotopy class of the geometrical $S^1$ factor (i.e., some
loop not homotopic to an integral curve of $\partial/\partial\phi$)
is contractible in $\tSi$. For further discussion of this issue see
Remark IV.3. We note that in $3+1$ dimensions, the condition
$\pi_1(\tSi)\approx \zz^{n-2}$ holds automatically, cf.\ \cite{tc}.

\noindent{\bf Proof.} There are two main parts to this proof.
First, using the splitting theorem and the topological assumption
(c), we show that the number of $\re$ factors in the splitting is
precisely $n-2$. This means that the submanifold $W$ is
$2$-dimensional. Using the topological censorship theorem
\cite{tc}, we then show that $\tSi\simeq T^{n-2}\times W$. The
only undetermined functions are then the 2-dimensional metric
$\ts_{AB}(\vec y)$ on $W$ and the lapse $\tN(\vec y)$, which, with
the aid of the field equations, can be solved for explicitly.

%\newpage

\noindent ({\it i}) \underbar{The product structure of $\tSi$:}

\noindent As discussed in (\ref{cc}), the assumption of negative mass
along with condition (S) implies condition (C). Thus, we can invoke
the splitting theorem which tells us that there are $k$ noncompact
directions in the universal covering space $\tSi^*$, so that
$\tSi^\ast = \re^k \times W$, where $W$ is compact.
By results on topological censorship \cite{tc}, the homomorphism
$i_* : \pi_1(\partial\widetilde \Sigma) \to
\pi_1(\widetilde \Sigma)$
is onto. But $\partial\widetilde \Sigma = A\times B$, where $A$ is
the $(n-2)$ torus and $B$ is the circle of assumption (a), whence
$\pi_1(\partial \tSi)\approx \pi_1(A)\times\pi_1(B)$.
Since, by assumption~(c), $\ker i_* = \pi_1(B)$, it follows that
$i_*|_{\pi_1(A)}:\pi_1(A)\to \pi_1(\widetilde \Sigma)$
is an isomorphism, and hence $\pi_1(\tSi) \approx \mathbb Z^{n-2}$.
But by Remark III.2.3, $\pi_1(\tSi)$ must contain a subgroup of finite
index isomorphic to $\mathbb Z^k$, from which it follows that $k = n-2$.

Thus, the universal
cover ${\tSi}^\ast$ splits isometrically as $\re^{n-2}\times W$.
$W$ is therefore a $2$-dimensional, compact, simply connected manifold
with boundary and is thus diffeomorphic to a disk.
The covering isometries of ${\tSi}^\ast$
split as $\fg=(\fg',\fg'')$, where $\fg'$ and
$\fg''$ are isometries of $\re^{n-2}$ and
$W$, respectively.
Let $A^*\simeq \re^{n-2}$ denote the universal cover of $A$. Then
$A^*\times B$ is the covering of $A\times B$ associated with the
subgroup $\pi_1(A)\subset \pi_1(A\times B)$. Since
$i_*|_{\pi_1(A)}:\pi_1(A)\to \pi_1(\widetilde \Sigma)$ is an isomorphism, the
group of covering isometries of ${\tSi}^\ast$ is naturally isomorphic
to $\pi_1(A)$. By this isomorphism, the covering isometries of ${\tSi}^\ast$
restrict on
$\pd \tSi^*\simeq \re^{n-2}\times S^1$ to the covering isometries of
$A^*\times B$. Since the latter fix the circle $B$, the former fix the disk
$W$, i.e., the covering isometries of ${\tSi}^\ast$ act only on the $\re^{n-2}$
factor. It follows that
\begin{equation}
\tSi \simeq (\re^{n-2}\times W)/\pi_1(A) \simeq (\re^{n-2}/ \pi_1(A))\times W
\simeq A\times W\simeq T^{n-2} \times D^2 \label{UT0},
\end{equation}
where $D^2$ is a closed 2-disc and $T^{n-2}$ is the $(n-2)$-torus of assumption
(a).

\noindent ({\it ii})$\,$\underbar{Solving for the lapse and metric on the
disc:}

\noindent
By the splitting obtained in part (i), the Fermat metric takes
the form
\begin{equation}
d\tilde s^2 = \sum\limits_{i=1}^{n-2} (du^i)^2 + {\ts}_{ij}(\vec
y) dy^idy^j,\quad \frac{\partial N}{\partial u^i}=0, \label{UT1}
\end{equation}
where $\vec y = (y^1,y^2)$ and $\ts_{ij}$ are the coordinates and
metric on the disk $W \simeq D^2$, respectively. Since there are
$n-2$ flat directions, the only contribution to the curvature comes
from $\bK^m{}_{ijk}$, the curvature tensor associated with
$\ts_{ij}$. As $\ts_{ij}$ is $2$-dimensional,
the curvature tensor can be expressed as, $\bK^m{}_{ijk}
=\frac{1}{2}{\bK}\,(\ts_{ik}\ts_j^{\,\,\, m} - \ts_i^{\, \, \,
m}\ts_{jk})$, where $K$ is the scalar curvature.
The field equations (\ref{FE4}) thus simplify to
\begin{equation}
 \bD_i\bD_j\tN= -\frac{\tN \bK}{2(n-1)} \ts_{ij}\quad ,
\label{UT2}
\end{equation}
where $\bD$ is the covariant derivative compatible with $\ts_{ij}$.
Differentiating, antisymmetrizing, and contracting
with $\ts^{kj}$, we get
\begin{equation}
\label{UT2.5}
\ts^{kj}\bD_{[k} \bD_{i]} \bD_j \tN =
-\frac{\ts^{kj}}{2(n-1)}\ts_{j[i}\bD_{k]}(\tN \bK)\quad ,
\end{equation}
which yields the integrability condition $\bD_i(\tN \bK) = (n-1) \bK
\bD_i (\tN)$. Thus, $\bK=C \tN^{n-2}$, where $C$ is an integration
constant.

Let us now consider a neighbourhood $\cU$ of the boundary
$\pd W\simeq S^1$. In this neighbourhood, we can write $\ts_{ij}$ in
Gaussian normal coordinates $(\rho, \phi)$ with respect to the
boundary at $\rho=0$ as
\begin{equation}
d\ts^2 = d\rho^2 + \fG^2(\rho,\phi)d\phi^2\quad.
\label{UT5}
\end{equation}
Using the condition $\bK=C\tN^{n-2}$,
equations (\ref{UT2}) become,
\begin{eqnarray}
\frac{\pd^2 \tN}{\pd \rho^2} &=& -\frac{C}{2(n-1)}
\tN^{n-1}\quad ,\label{UT6}\\ \frac{\pd^2 \tN}{\pd \rho \pd \phi} &=&
\frac{1}{\fG}\frac{\pd \fG}{\pd \rho} \frac{\pd \tN}{\pd
\phi}\quad ,\label{UT7}\\ \frac{\pd^2 \tN}{\pd \phi^2} &=& -\fG \frac{\pd
\fG}{\pd \rho}\frac{\pd \tN}{\pd \rho} +\frac{1}{\fG}\frac{\pd
\fG}{\pd \phi}\frac{\pd \tN}{\pd \phi} -\frac{C\tN^{n-1}\fG^2}{2(n-1)}
\quad .\label{UT8}
\end{eqnarray}

The boundary conditions for this system of equations at $\rho = 0$ are
\begin{equation}
\fG(0,\phi)=\frac{1}{\ell}, \quad \frac{\pd \tN}{\pd
\phi}\biggl \vert_{\rho=0}=0\quad , \quad
\frac{\pd \tN}{\pd \rho}\biggl \vert_{\rho=0}=\frac{1}{\ell}\quad .\label{UT9}
\end{equation}
The first of these conditions arises by comparison of (\ref{UT5}) and
assumption (a) of the Theorem, the second arises since we assume $\tN=0$
pointwise on the boundary, and the last is obtained by combining
(\ref{FE6}), (\ref{FE7}), (\ref{BC15}), (\ref{UT5}), and the second
boundary condition. Now, (\ref{UT7}) simplifies to
\begin{equation}
\pd_\rho \left ( \frac{1}{G}\pd_\phi\tN\right ) =0 \quad \Rightarrow \quad
\frac{1}{G}\pd_\phi\tN =\xi(\phi)
\quad . \label{UT10}
\end{equation}
(\ref{UT9}) implies that $\xi(\phi)=0$, or
$\pd_\phi \tN=0$ everywhere. Thus, (\ref{UT6}) reduces to an
ordinary differential equation,
\begin{equation}
\frac{d^2 \tN}{d \rho^2} = -\frac{C}{2(n-1)}\tN^{n-1}\quad,
\label{UT11}
\end{equation}
which integrates to give
\begin{equation}
\frac{d\tN}{d\rho}=
\frac{1}{\ell}\left(1-\frac{C\ell^2\tN^n}{n(n-1)}\right)^{1/2}\quad ,
\label{UT12}
\end{equation}
given the boundary conditions (\ref{UT9}). Next, combining
(\ref{UT6}) and (\ref{UT8}) and using $\pd_\phi \tN=0$, we get
\begin{equation}
\pd_\rho \left (\frac{1}{G}\pd_\rho \tN\right )= 0 \quad \Rightarrow \quad
\frac{1}{G}\pd_\rho \tN =
\zeta(\phi)\quad , \label{UT13}
\end{equation}
which along with (\ref{UT9}) implies that
$\zeta(\phi)=1$, or $\pd_\phi\fG=0$ identically, so that
$\fG(\rho)=\frac{d\tN}{d\rho}$.

Using (\ref{UT12}) we can rewrite the metric (\ref{UT5}) as
\begin{equation}
d\ts^2 = \frac{\ell^2}{r^4 \left(1-\frac{C\ell^2}{n(n-1)r^n}\right)}
dr^2 + \frac{1}{\ell^2} \left(1-\frac{C\ell^2}{n(n-1)r^n}\right)d\phi^2
\quad ,
\quad
\tN=\frac{1}{r}=\Omega\quad . \label{UT14}
\end{equation}
Inserting into (\ref{UT1}) and comparing with
(\ref{Intro1}), we see that the particular identification
$\frac{C\ell^{2}}{n(n-1)}=r_0^n$ yields precisely the soliton metric
in the neighbourhood $\cU$ of the boundary.

Finally, we note that the assumption of
negative mass implies that $r_0>0$ and therefore $C>0$.
Our coordinates, valid on a neighbourhood of scri, thus extend to all
$r>r_+^n:=\frac{C\ell^{2}}{n(n-1)}$, and the metric is unique on this
domain. Now, by the boundary conditions, $\phi$ is periodic with period
$\frac{4\pi \ell^2}{nr_0}$, and so the further extension to $r=r_+$ is
a cone point singularity, except that it is a smooth point iff
we choose $C$ such that $C=C_0:=\frac{n(n-1)r_0^n}{\ell^2}$ (as can
be seen by the standard technique of expanding $d\ts^2$ in polar
coordinates about $r=r_+$).
Thus, there is a unique nonsingular extension of the solution in the
neighbourhood $\re^{n-2}\times \cU$ of scri to all of
$\re^{n-2}\times W \simeq \re^{n-2} \times D^2$.\qed

\noindent {\bf Remark IV.2.} One may impose in Theorem II1.2.1 and,
subsequently, Theorem IV.1 a weaker regularity requirement, i.e.,
one may assume that $(\Sigma, h, N)$ is $C^k$
conformally compactifiable, for some $k\ge n+1$, but then
the isometries constructed will in general be of finite
differentiability.

\noindent {\bf Remark IV.3 (Non-Uniqueness).} Let us consider, in
Theorem IV.1, the somewhat more general situation in which $\pd
\tSi$ is an $(n-1)$-torus with a fixed, but arbitrary, flat
structure. Each free homotopy class of $\pd \tSi$ can be
represented by a closed geodesic $\b$.  Suppose more generally, in
Theorem IV.1, that a fixed but arbitrary embedded closed geodesic
$\b$  generates the kernel of $i_*$.  One still has in this
somewhat more general setting that $\pi_1(\tSi)={\Bbb Z}^{n-2}$.
As in the proof of Theorem IV.1, it follows that the universal
cover of $\tSi$ is isometric to Euclidean $(n-2)$-space times a
$2$-disk, whose boundary circle has length $L(\b)=$ length of
$\b$. One can again solve for the geometry on the $2$-disk to
conclude that the universal cover of spacetime is isometric to an
{\it open} adS soliton (spacetime with metric of the form I.1, but
where the $\theta^i$ coordinates are no longer periodic but range
through all of ${\Bbb R}$). Thus, the physical spacetime $M$ is
isometric to an open adS soliton quotiented out by a certain $\Bbb
Z^{n-2}$ action, $M\simeq$ (open AdS Soliton)$/\Bbb Z^{n-2}$.
Since the cycle $\b$ may no longer be {\it geometrically} an $S^1$
factor of $\pd\tSi$, the $Z^{n-2}$-action may no longer fix the
disk $W$, as it had done in part ({\it i}) of the above proof, and
we may lose the global geometric factorization (\ref{UT0}). Nonetheless,
the relevant  actions have been described explicitly in the $3+1$
(i.e., $n=3$) case in \cite{Anderson}. In this case one obtains a
countable class of locally isometric but isometrically distinct
spacetimes, {\it generalized} AdS solitons, whose mass depends on
$L(\b)$. Similar behavior will occur in higher dimensions.

Thus, fixing the conformal boundary, we obtain a spectrum of
solitons of differing masses as we vary the choice of cycle $C$
representing ${\rm ker}\,i_*$. For a given flat torus, the ground
state soliton arises by choosing $C$ to be the shortest cycle on
the torus. Other choices give rise to solitons that are excited
states (they are still {\emph local} minima of energy, by the
Constable-Myers analysis \cite{cm}). The excited states have roles
to play in the AdS/CFT correspondence. For example, they
contribute to the Conformal Field Theory partition function
\cite{DMMV}. A particularly interesting application is to consider
homotopically distinct closed geodesics $g_1,g_2,g_3\dots$ on a
given torus, ordered by increasing length. Consider the solitons
$S_1$, $S_2$ that arise by taking $C_1$, respectively $C_2$, as
the bulk-contractible cycle $C$. Then $S_1$ is the ground state
and $S_2$ is the first excited state. Now vary the conformal
structure so that the lengths of $g_1$ and $g_2$ cross, but remain
less than the lengths of the other homotopically distinct cycles,
so (the variation of) $S_2$ becomes the ground state. Page
\cite{Page} has shown that this produces a zero temperature phase
transition in the boundary CFT, as the pressures exerted on area
elements transverse to $g_1$ and $g_2$ suddenly change sign.

\section{Concluding Remarks}\label{concluding.sec}
\setcounter{equation}{0}

\noindent In a sequel to this paper
\cite{GSW2}, we make use of many of the techniques and results
presented here to study static, asymptotically locally AdS
spacetimes which contain black hole horizons (e.g., which contain
non-naked singularities). For instance we show that a static
spacetime asymptotic to the AdS soliton cannot have negative mass
if it contains a horizon. This result is well illustrated by the
toroidal Kottler spacetimes. In the sequel, we will also apply the
results of Section II to study the case of an asymptotically
locally AdS spacetime with conformal boundary admitting
non-negative Ricci curvature.

\section*{Acknowledgements}
\setcounter{equation}{0}

\noindent
This work was partially supported by grants from the National Science
Foundation (USA) DMS-0104042 and the Natural Sciences and Engineering
Research Council (Canada). SS was supported by a postdoctoral fellowship
from the Pacific Institute for the Mathematical Sciences. GJG wishes to
thank DAMTP, University of Cambridge, for hospitality during the completion
of part of this work.

\section*{Note Added}
\setcounter{equation}{0}

\noindent
After this paper was submitted, \cite{ACD} appeared,
containing related results in the $4$-dimensional case.

\vfill\par\eject

\appendix
\section{Hawking-Horowitz Mass} \label{appendix.sec}
\renewcommand{\theequation}{\thesection.\arabic{equation}}
\setcounter{equation}{0}

\noindent Under the conformal transformation that replaces the geometry
induced by the Fermat metric with that induced by the spacetime
metric, the respective mean curvatures obey
\begin{eqnarray}
H&=&x\tH-(n-1)\tn^c\tnabla_cx\nonumber\\
&=&x\tH+\frac{(n-1)}{\psi}\quad .\label{A1}
\end{eqnarray}
The $x\tH$ term can be estimated using (\ref{BC20}), but to estimate
the second term requires some further work. We begin with the
identity (\ref{FE7}), which together with the definition $\tW=1/\psi^2$
gives
\begin{equation}
\frac{1}{\psi^2}=\frac{-2\Lambda-x^2\tR}{n(n-1)}\quad .\label{A2}
\end{equation}
Now this can be estimated by combining
(\ref{FE11}), (\ref{BC22}), and the estimates of Section II to obtain
\begin{equation}
\tR=\frac{x}{n-2}\frac{\partial \tR}{\partial x}+{\cal O}(x^n)
=\frac{x^{n-2}}{(n-2)!}\tR^{(n-2)}(0)
+{\cal O}(x^n)\quad ,\label{A3}
\end{equation}
which can be substituted into (\ref{A2}) to yield
\begin{equation}
\frac{1}{\psi^2}=\frac{-2\Lambda}{n(n-1)}-\frac{x^n}{n!}
\tR^{(n-2)}(0)+{\cal O}(x^{n+2})\quad .\label{A4}
\end{equation}
We take the square root, substitute the result into the last term in
(\ref{A1}), and use (\ref{BC20}) to estimate the first term, obtaining
\begin{equation}
H=\frac{(n-1)}{\ell}+\frac{x^n\ell}{2(n!)}\tR^{(n-2)}(0)
+{\cal O}(x^{n+2})\quad .\label{A5}
\end{equation}

Following Hawking and Horowitz, we multiply $H$ by the lapse
$N=1/x$ and integrate over a constant $x$ surface $Y_x$ using the
volume element $\sqrt{b}=\sqrt{\tb}/x^{n-1}$ of the (unrescaled)
first fundamental form $b_{ab}:=N^2\tb_{ab}$. This gives
\begin{equation}
\int_{Y_x}NH\sqrt{b}\ dY_x = \frac{(n-1)}{x^n\ell} {\rm
vol}(Y_x) +\frac{\ell}{2(n!)}\int_{Y_x}
\tR^{(n-2)}(0)\sqrt{\tb}\ dY_x +{\cal O}(x^2) \quad .\label{A6}
\end{equation}

By the Hawking-Horowitz prescription, we must now embed a surface
$Y_x'$ in a slice of a reference spacetime, compute its mean
curvature $H_0$, pull this back to $Y_x$ (call the pullback
$H_0^*$), integrate it as above, compute the difference between
the resulting integral and (\ref{A6}), and finally take $x\to 0$.
The intrinsic geometry on $Y_x'$ induced from the reference
spacetime must match that on $Y_x$. The reference spacetime we
will use is of constant negative curvature and is assumed to have
the same Penrose conformal boundary as the physical spacetime
(this may force the reference spacetime to be incomplete; see
\cite{hm} for a discussion). We embed within a moment of time
symmetry thereof, so the scalar curvature of this slice is
constant. Thus
\begin{equation}
H_0=\frac{(n-1)}{\ell}+{\cal O}(x^{n+2}) \quad ,\label{A7}
\end{equation}
and thus
\begin{equation}
\lim_{x\to 0}\int_{Y_x}N(H-H_0^*)\sqrt{b}\ dY_x
=\frac{\ell}{2(n!)}\int_{Y}\tR^{(n-2)}(0)\sqrt{\tb}\ dY
\quad ,\label{A8}
\end{equation}
where $Y:=\lim_{x\to 0}Y_x$. Thus we have:

\noindent {\bf Proposition A.1.} {\sl Let $(\Sigma,h,N)$ obey equations
(\ref{FE2}--\ref{FE3}) and be $C^n$ conformally compactifiable.
Then the Hawking-Horowitz mass $M_{HH}$ of $\Sigma$ is given by}
\begin{equation}
M_{HH}=\frac{-\ell}{16\pi(n!)} {\int_Y \tR^{(n-2)}(0)\sqrt{\tb}\ dY}
\quad . \label{A9}
\end{equation}

\noindent{\bf Proof.} Follows from (\ref{A8}) by comparing its
left-hand side to the definition of $M_{HH}$ appearing in equation
(2.14) of \cite{HH}. \qed

\noindent {\bf Remark A.2.} By setting $Y=\partial \tSigma$ and
comparing (\ref{A9}) to (\ref{BC35}), we have $M_{HH}=M_{AM}$ in
this setting.

\section{Fermat and Null Lines}
\setcounter{equation}{0}

\noindent {\bf Lemma B.1.}
Let $(\Sigma, h, N)$ be a static spacetime with lapse $N$ bounded from
below. Let $\pi:(M,g_{ab})\to (\tSigma,\wh_{ab})$ be the projection
along integral curves of the timelike
Killing field $\partial/\partial t$. Then, for every Fermat line
$\rho:{\mathbb R}\to(\tSigma,\wh_{ab})$ through $p$
there is a unique future directed null line $\gamma:{\mathbb R}\to
(M,g_{ab})$ through $p$, with $\rho=\pi(\gamma)$.

\noindent
{\bf Remark.} There are in fact two null lines through $p$ that project
to $\rho$, but one reverses orientation.

\noindent
{\bf Proof.} Let $\ts$ be an arclength (i.e., affine) parameter
along $\rho$. Without loss of generality, we can take $\rho$ to be of
unit speed. On $\pi^{-1}(\rho)$, we can thus define the
future directed null vector field, $\frac{\partial}{\partial t} + {\dot
\rho}(\ts)$ and let $\gamma$ be the unique integral curve
through $p$. Then along $\gamma$, we have $dt=d\ts$.

We now show that $\gamma$ is achronal. Let us
assume otherwise. Then there exist points $q,q' \in \gamma$, with
$q=\gamma(a)$, $q'=\gamma(b)$ and $a<b \in \mathbb R$, such that there
is a timelike curve $\gamma'$ from $q$ to $q'$. Let $\rho'$ be its
projection into $\Sigma$. Since $dt=d\ts$ along $\gamma$, we can
integrate this to obtain $\Delta t:=b-a =\tL(\rho)$, which is the Fermat
length of $\rho$ between the projected points $\pi(q)$ and
$\pi(q')$. Since $\rho$ is a Fermat line, $\tL(\rho)$ equals the Fermat
distance from $\pi(q)$ to $\pi(q')$. Using $\ts$ to also parameterise
$\gamma'$, we see that $dt > d\ts$,
which integrates to give $\Delta t=b-a> \tL(\rho')$, the length of
$\rho'$ between $\pi(q)$ and $\pi(q')$. But this means that $L(\rho')
< L(\rho)$, which contradicts the claim that $\rho$ is minimal.

Finally, we show that $\gamma$ is also complete. That it is geodesic
follows from the fact that it is null and achronal. However, the
parameterisation $\ts$ is not affine. Using the condition of staticity
it can be shown
that, up to a constant of proportionality, $\gamma$ has affine
parameter $\lambda$ related to $\ts$ by $d\lambda=N^2 dt=N^2d\ts.$
Now, assume that $\gamma$ is incomplete, i.e., that it is inextendible
in at least one direction in which it has a bounded range of affine
parameter. Since $N$ is bounded away from zero, this means that the
affine parameter $\ts$ in that direction is also
bounded above, so that $\rho$ would be incomplete,
which again is a contradiction.

Thus, $\gamma$ is a complete achronal null line. \qed

\end{document}